\documentclass[preprint,authoryear,12pt]{elsarticle}

\RequirePackage{amsmath}
\RequirePackage{amsfonts}
\RequirePackage{amssymb}
\RequirePackage{graphicx}
\RequirePackage{subfigure}
\RequirePackage{natbib}

\RequirePackage{times}
\RequirePackage{mathptm}

\graphicspath{{Figs/}} 

\let\vec\undefined 
\let\mat\undefined %

\newcommand{\mat}[1]{\mathbf{#1}} 
\newcommand{\vec}[1]{\boldsymbol{#1}} 

\newcommand{\by}{ {\boldsymbol y} }

\newcommand{\chisq}{\chi^2}

\newcommand{\tX}{\widetilde{\mat{X}}}
\newcommand{\ty}{\widetilde{\vec{y}}}

\newcommand{\hb}{\widehat{\vec{\beta}}}

\journal{Remote Sensing of Environment}

\begin{document}

\begin{frontmatter}
  
\title{Bayesian principal component regression model with spatial effects for forest inventory variables under small field sample size}

\author[lut]{Virpi Junttila\corref{cor}}
\ead{virpi.junttila@lut.fi}

\author[fmi,lut]{Marko Laine}
\ead{marko.laine@fmi.fi}

\address[lut]{Lappeenranta University of Technology, School of Engineering Sciences,
P.O. Box 20, FI-53851 Lappeenranta, Finland}
\address[fmi]{Finnish Meteorological Institute, P.O. Box 503, FI-00101 Helsinki, Finland}

\cortext[cor]{Corresponding author}


\begin{abstract}
Remote sensing observations are extensively used for analysis of environmental variables. These variables often exhibit spatial correlation, which has to be accounted for in the calibration models used in predictions, either by direct modelling of the dependencies or by allowing for spatially correlated stochastic effects. 
Another feature in many remote sensing instruments is that the derived predictor variables are highly correlated, which can lead to unnecessary model over-training and at worst, singularities in the estimates.
Both of these affect the prediction accuracy, especially when the training set for model calibration is small. To overcome these modelling challenges, we present a general model calibration procedure for remotely sensed data and apply it to airborne laser scanning data for forest inventory. We use a linear regression model that accounts for multicollinearity in the predictors by principal components and Bayesian regularization. It has a spatial random effect component for the spatial correlations that are not explained by a simple linear model. An efficient Markov chain Monte Carlo sampling scheme is used to account for the uncertainty in all the model parameters. We tested the proposed model against several alternatives and it outperformed the other linear calibration models, especially when there were spatial effects, multicollinearity and the training set size was small.
\end{abstract}

\end{frontmatter}


\section{Introduction}
\label{sec:intro}

Remotely sensed data, e.g.\ from satellites, digital aerial images, or airborne laser scanning, are increasingly used for mapping ecological variables over large geographical areas. Typical examples of such use are habitat and biodiversity monitoring \citep{mcdermid09,nagendra13}, lake water quality \citep{matthews10} and forest inventory \citep{masek15}. Remotely sensed observations provide only indirect information of the area and modelling is needed for the interpretation of the data in terms of the variables of interest. In forest inventory, these variables include forest characteristics such as average forest biomass, median tree height, per hectare stem number or average timber volume, see e.g. \cite{naesset:1997}, \cite{means:1999}, \cite{RookerJensen:2006}, and \cite{magnussen:2010}. In some cases, direct physical models might be available (e.g.\ based on emission and scattering of light), but generally a simpler black-box type model to translate the remotely sensed data to the ecological variables is needed, e.g. linear regression (see the references above) or other methods \citep[see e.g.][]{powell10,gleason12,belgiu16}. In forest inventory, this is typically done by linear regression.

A benefit of remotely sensed observations is that they can cover the whole spatial area of interest and the geographically located ecological variable can be predicted over the whole area in a pixel or some other sub-area level. To calibrate the model for prediction, i.e., to estimate the model parameters, remotely sensed data need to be accompanied by a set of field measurements of the ecological variables at chosen test locations. For instance, in area based prediction of forest inventory variables, the field measurements can be given as per hectare values estimated in circular field sample plots with a given radius. In forest inventory models, the number of field sample plots needed for accurate calibration can be several hundreds for an area between 10,000 and 100,000 hectares \citep{maltamo:2011}.
The design of the field measurement locations has to account for not only the obvious statistical properties, but also the landscape properties such as mountainous terrain or thick forest, and it may be laborious and costly to reach these locations for the measurement work. Thus, to decrease the cost of the predictions, 
it is preferable to keep the number of field measurements to a minimum.
This ambition for a small training set causes additional challenges to the model parameter estimation process since the regression problem may become under-determined and easily suffer from the effects of over-training \citep{junttila13}. 

LiDAR (Light Detection and Ranging) is an active remote sensing system based on laser light. In airborne LiDAR, a sensor in an aeroplane or a helicopter sends laser pulses towards the ground and records the time lapses between the launch of the beams and the return of the signals. In area based models, the LiDAR predictor variables are usually some statistical aggregates of the actual LiDAR pulse measurements over the geographical sub-areas. These variables are typically highly correlated and this multicollinearity can cause singularities in the model. 
{If the number of LiDAR predictor variables is large compared to number of field sample plots, the multicollinearity can also cause unnecessary model over-training because, in some sense, the highly correlated predictor variables contain the same information about the response and thus add no new information, only redundant variables.}
 A general approach to overcome problems caused by multicollinearity is to use variable selection algorithms or principal component regression. In a recent study, \citet{junttila14} achieved good results with a small training set and highly correlated multidimensional data by utilizing singular value decomposition combined with Bayesian regularization.

Many ecological variables are spatially correlated, which means that experimental units geographically close to each other are likely to be more similar than those far away. If the model explains this variability well, the model based predictions follow the same correlation. However, any lack-of-fit, which is inevitable in most linear calibration models, may produce spatially correlated model residuals, i.e., residuals geographically close to each other are more similar than those far away by residual sign and amplitude. In such occasions, the model performance is improved if the predictions are corrected toward those field measurements close to the prediction location. 

In this paper, we build a linear model for prediction of the ecological variable of interest with a small number of field measurements which is both efficient and robust against modelling assumptions.
We use a Bayesian approach that allows us to implement effective estimation of complex, hierarchically structured parameter associations appropriate for accommodating the strong multicollinearity typical in remotely sensed predictors and spatial autocorrelation among the model residuals.
The method is applicable for many problems where strongly correlated data are translated to spatial observations with a linear model.
In the proposed model, the problems caused by the multicollinearity of the predictors and by the small number of field measurements are overcome by using predictor orthonormalization and regularization. We utilize Bayesian regularization to emphasize those linear combinations of principal component predictors that explain most of the variability of the original predictor variables and that have predictive information on the ecological variable of interest. A general spatial dependency is allowed for the residuals of the model by a spatial random effect. 
The hierarchical model is estimated and uncertainty in the spatial model parameters is 
carried through to the predictions by using an efficient adaptive Metropolis Markov chain Monte Carlo (MCMC) algorithm.

We validate model performance by using both synthetic data with different noise levels and spatial correlation, and with real-world observations for forest inventory. In both cases, we assume a given design for the field plot locations and show how the spatial correlation structure and a full Bayesian treatment of model parameter uncertainties improve the model based predictions. 

This work is based on earlier studies by \citet{junttila13}, who used a similar spatial model, but with plug-in estimates of the spatial parameters instead of MCMC, and studies by \citet{junttila14}, who used combination of singular value decomposition and regularization for regression parameters, but solved them with maximum likelihood estimation instead of MCMC. The parametric uncertainties of the two earlier papers are now dealt with using a sampling based approach. We show, by comparing the predictive power, that the approach chosen here outperforms the earlier methods in the presence of spatial correlation in the model error.

The article is organized as follows: we first define the proposed model in Section~\ref{sec:materialsandmethods}; the used datasets, both synthetic and real, are described and the validation procedure is explained in Section~\ref{sec:testdata}; the results of the validation are given in Section~\ref{sec:results}; and finally conclusions based on the results are given in Section~\ref{sec:discussion}.

\section{Statistical Methods}
\label{sec:materialsandmethods}

\subsection{The proposed model}
\label{sec:stat}

In this study, we use a linear regression model with a spatial random effect and hierarchical shrinkage prior for the regression parameters. The model combines the spatial modelling and singular value decomposition regularization described in \citet{junttila13} and \citet{junttila14}. Instead of maximum a posteriori (MAP) estimates, the model parameters are estimated using Markov chain Monte Carlo (MCMC) simulation.
With MCMC we can implement a hierarchical Bayesian model that can handle complex structured parameter associations and  fully account for the uncertainty in all the model parameters for the model based predictions of the ecological variables of interest outside the training set.

We write our model as
\begin{equation}
  \vec{y} = \mat{X}\vec{\beta} + \vec{\eta} + \vec{\epsilon},\quad
  \vec{\eta} \sim N(\vec{0},\mat{C}), \quad
  \vec{\epsilon} \sim N(\vec{0},{\tau}^2 \mat{I}),
  \label{eq:linmodel}
\end{equation}
where {the response $\vec{y}$ is a vector of the ecological variable of interest containing $n$ observations, $\mat{X}$ is an $n\times (p+1)$ matrix that includes an intercept column and $p$ columns of principal components of the remotely sensed data based variables, as described in Section~\ref{sec:pca},} $\vec{\beta}$ is a $p+1$ vector of regression parameters, $\vec{\eta}$ is an $n$ vector for the spatial random effect, and $\vec{\epsilon}$ is an $n$ vector for non-spatial error. Both the random terms, $\vec{\eta}$ and $\vec{\epsilon}$, are assumed Gaussian. A full covariance matrix $\mat{C}$ defines the spatial correlation structure of the model residuals by using distances between the field measurement locations. The errors in $\vec{\epsilon}$ are assumed independent and identically distributed with variance $\tau^2$. 

Data at $n$ locations containing the field measurements of the ecological variable are referred as the training set, while the locations where the variable needs to be predicted, are referred as the validation set. The predictors and the geographical coordinates are assumed to be known in each training set and validation set location. 

To estimate the model parameters, we use hierarchical formulation to define the priors. 
{To obtain the predictor regularization effect using the priors, we follow the formulation of \cite{tipping:2001}.} For the regression parameter ${\beta}_i,\ i = 0,1,2,\ldots,p$, the prior is zero mean Gaussian with inverse of variance $\alpha_i$.
The variance parameters ${\alpha}_i$ are assumed to be unknown and they are estimated too. The prior for ${\alpha}_i$ is defined by using a scaled $\chi^2$ distribution and we have
\begin{align}
  \beta_i    &\sim N(0,\alpha_i^{-1}), \quad i=0,1,\ldots,p, \label{eq:betapri}\\
  \alpha_i   &\sim \chisq(\nu_{i},a_{i}), \quad i=0,1,\ldots,p.\label{eq:alphapri}
\end{align}
The scaled $\chi^2$ distribution is common in Bayesian analyses and can be defined by the standard Gamma distribution as $\chisq(\nu,a) = \Gamma(\nu/2,a\nu/2)$ \citep[see, e.g.][]{gelman03}. The scaled $\chi^2$ parameterization is convenient in applications. We can interpret it as if knowing the value of ${\alpha}$ to be ${a}$ from ${\nu}$ (virtual) previous observations.  In addition, it is the conjugate distribution for the inverse variance, which allows for the Gibbs sampling approach outlined below. 

Here, we follow \citet{junttila14} and use common inverse variance $\vec{\alpha}$ prior for all regression parameters $\vec{\beta}$. This will shrink the non-significant parameters to zero in correct proportions and the amount of this shrinkage (the value of $\vec{\alpha}$) will be determined from the data by the estimation procedure and thus no arbitrary truncation of the number of used components is needed.
We use a separate prior, $\alpha_0$, for the intercept parameter of the model (\ref{eq:linmodel}), i.e.\ for the first component of $\vec{\beta}$, $\beta_0$, corresponding to the first column of $\mat{X}$, and a common prior for all the rest of the components of $\vec{\beta}$, $\alpha_i = \alpha_1$ for $i=1,2,\ldots,p$. Thus, we define the $p$ vector with two distinct parameter values $\alpha_0$ and $\alpha_1$, $\vec{\alpha} = [\alpha_0,\alpha_1,\alpha_1,\dots,\alpha_1]^T$, and similarly for the prior parameters $\vec{a}$ and $\vec{\nu}$.
Instead of maximum a posteriori (MAP) approach for the solution of the parameters used in \cite{tipping:2001} and \cite{junttila14}, we use a simulation based procedure that estimates the linear model parameters as well as the spatial parameters by their full posterior distributions. 

The covariance matrix $\mat{C}$ in Eq.~(\ref{eq:linmodel}) describes the spatial dependence between the model residuals after accounting for the linear effects in $\mat{X}\vec{\beta}$. We assume a simple parametric relationship and use a convenient exponential covariance function defined by a correlation decay parameter $\phi$ and a spatial variance parameter $\sigma^2$ as 
\begin{equation}
  \label{equ:sigmaeta}
  \mat{C}_{ij} = \sigma^2 e^{-d(i,j)/\phi},
\end{equation}
where $d(i,j)$ is a geographical distance of points between two field measurement locations $i$ and $j$ \citep{gelfand10}.
We define the total covariance matrix for the observations as $\vec{\Sigma}_\theta = \mat{C} + {\tau}^2 \mat{I}$, which depends on a parameter vector $\vec{\theta} = [\tau^2,\sigma^2,\phi]^T$. To complete the description, we need a prior for $\vec{\theta}$,  for which we have used log-normal distribution, 
\begin{equation}
  \label{equ:thetapri}
{\theta}_j\sim \log N({\mu}_{\theta j},{\sigma}^2_{\theta j}),\quad j=1,2,3,
\end{equation} 
independently for each component.
This is a conventional choice, and we choose the prior hyper parameters to reflect our assumptions on the spatial structure.
As we will be using a Metropolis-Hastings sampling step for $\vec{\theta}$, we are not restricted to conjugate priors and could use any parametric prior distribution.

Equation (\ref{equ:sigmaeta}) defines the covariance between any two locations at distance $d$, not just for those where we have observations. As an approximation, the correlation is assumed to depend on the distance only. It would be possible to construct a covariance function that takes into account the geography in more detail also. The chosen parametric form of the spatial covariance did perform well in our real world application.

\subsection{Computational procedure}
\label{sec:comp}

Given the values of parameters $\vec{\alpha}$ and $\vec{\theta}$, which define the regression hyper parameters and the covariance structure, the model in equation (\ref{eq:linmodel}) is a linear regression model with Gaussian errors. Therefore the conditional posterior distribution for the regression parameter $\vec{\beta}$ is known analytically. However, for full uncertainty quantification that accounts for all of the unknowns, we perform Markov chain Monte Carlo (MCMC) sampling for the joint distribution $p(\vec{\beta},\vec{\alpha},\vec{\theta}|\vec{y})$. This is achieved by the following steps that combine Gibbs sampling for $\vec{\alpha}$ and $\vec{\beta}$, and adaptive Metropolis-Hastings sampling for $\vec{\theta}$. 
By using standard arguments \citep[e.g.][]{gelman03}, we see that the conditional distribution $p(\vec{\beta}|\vec{y},\vec{\alpha},\vec{\theta})$ is Gaussian and we can sample from it directly. Similarly, as the scaled $\chi^2$ distribution is a conjugate distribution for the inverse variance in Gaussian error model, the distribution $p(\vec{\alpha}|\vec{y},\vec{\beta},\vec{\theta})$ is scaled  $\chi^2$, independently for each component, and the one dimensional marginal distributions can be sampled easily by standard algorithms \citep[e.g.][]{marsaglia00}. For the covariance parameters in $\vec{\theta}$, the closed form of the conditional distribution is not known and we perform an adaptive Metropolis-Hastings simulation for it \citep{dram}. The MCMC simulation will produce a chain of values of the parameters, which can be used as a sample from the full joint distribution of unknowns $p(\vec{\beta},\vec{\alpha},\vec{\theta}|\vec{y})$. With the chain we can estimate the joint and marginal distributions of the parameters and any distribution characteristics such as the mean. Furthermore, it can be used to assess the uncertainty in model based predictions of field variables at new field locations, where we have LiDAR observations but not the actual measurements used in calibration of the model. Further details on computations are given in \ref{sec:Acomp}.

\subsection{Principal components}
\label{sec:pca}

Typically, the predictor variables based on remotely sensed data can be highly correlated.  Following \citet{junttila14}, we use singular value decomposition of the predictors, and the analysis is performed using the principal components \citep{bair06}. The hierarchical model regularizes the problem and no arbitrary truncation of the number of principal components is needed.

Let the matrix $\mat{Z}$ contain centered and scaled transformations of the $p$ LiDAR predictor variables (i.e.\ each column has mean zero and standard deviation one). The matrix $\mat{Z}$ contains all the available values of the study area, including both the training set and validation set. Next, we do singular value decomposition,
\begin{equation}
  \label{eq:svd}
  \mat{Z} = \mat{U}\mat{S}\mat{V}^T,
\end{equation}
where $\mat{U}$ and $\mat{V}$ are orthonormal matrices of the left and right singular vectors, and the matrix $\mat{S}$ is a diagonal matrix of the singular values. Let $\mat{Z}_n$ be a matrix of those rows of $\mat{Z}$ where we have field measurements of the ecological variable (the training set). We form the predictor matrix $\mat{X}$ by multiplying $\mat{Z}_n$ by $\mat{V}$, and adding an intercept column of ones $\mat{1}_n$,
\begin{equation}
  \label{eq:x}
  \mat{X} =
  \begin{bmatrix}
    \mat{1}_n & \mat{Z}_n\mat{V}
  \end{bmatrix}.
\end{equation}

The columns of $\mat{X}$ (other than the first intercept column) are now ordered decreasing with respect to their variance, which allows variable selection type regularization using a hierarchical prior. The columns are only nearly orthogonal, as the principal components are evaluated from the full data of both the training set and the validation set, but the predictor matrix $\mat{X}$ uses only the subset referring to the training set.

As usual in principal component regression, we could use only the first $k$ components, and columns of $\mat{V}$, in the transformation, as typically the rightmost columns do not contain much information on variability in $\vec{y}$. Different versions of principal component regression analysis can be performed depending on how many principal component variables (columns of $\mat{X}$) are used and how we define the prior distributions for the regression parameters $\vec{\beta}$. 
{Here, we use common inverse prior $\alpha_1$ for all regression parameters associated with the principal components. The level of shrinkage for each principal component variable is determined by its singular value and by the inverse variance estimated from the data. Thus no arbitrary truncation of the number of used components is needed.} 
However, to discard the columns which do not explain the variability at all, and to avoid numerical problems, we neglect those predictors whose singular values are close to zero within the numerical accuracy.

\subsection{Prediction}

After we have estimated the model parameters, we can predict the value of the ecological variable, $y_\mathrm{new}$, at a new location where we have remotely sensed data but not actual field measurements. If $\vec{z}_\mathrm{new}$ contains the LiDAR predictor variable values scaled with the same scaling as the original $\mat{Z}$ in equation~(\ref{eq:svd}), we must transform it to the principal component coordinates used in the model estimation to obtain $\vec{x}_\mathrm{new}$,
\begin{equation}
  \label{eq:xnew}
  \vec{x}_\mathrm{new}  =
  \begin{bmatrix}
    1 & \vec{z}_\mathrm{new}\mat{V}
  \end{bmatrix}.
\end{equation}
Then, given values for the model parameters $\vec{\beta}$ and $\vec{\theta}$,  we can compute
\begin{equation}
  \label{eq:ynew}
    y_\mathrm{new} = \vec{x}_\mathrm{new}{\vec{\beta}} +
           \mat{C}_{\mathrm{new}}\vec{\Sigma}^{-1}_\theta(\vec{y}-\mat{X}\vec{\beta}),
\end{equation}
where $\mat{C}_{\mathrm{new}}$ is a row matrix that contains the spatial covariances between the new location and the original locations \citep{gelfand10}.
This calculation can be performed using point estimates of the parameters $\vec{\beta}$ and $\vec{\theta}$, only, or by repeating the calculation of $y_\mathrm{new}$ while sampling from the posterior distributions of the parameters to produce a predictive distribution for the new value. The latter approach is used here to sample from the predictive posterior distribution of $y_\mathrm{new}$ which accounts for the uncertainty arising from the estimation of $\vec{\beta}$ and $\vec{\theta}$.

\section{Materials and validation procedure} \label{sec:testdata}

We validate the model performance using two sets of data: synthetic data and real data from forest inventory.  Synthetic data are useful for the validation of the computational procedures and for comparing the different methods as we know the truth and can perform repeated sampling and analysis. The real life dataset demonstrates that the proposed model offers added benefits in forest inventory and it has potential for other similar modelling situations.

\subsection{Synthetic data}\label{test:toy}

\begin{figure}
  \centering
  \subfigure[$\vec{\eta}+ \vec{\epsilon}$]{\label{nue}
  \includegraphics[width=0.48\textwidth]{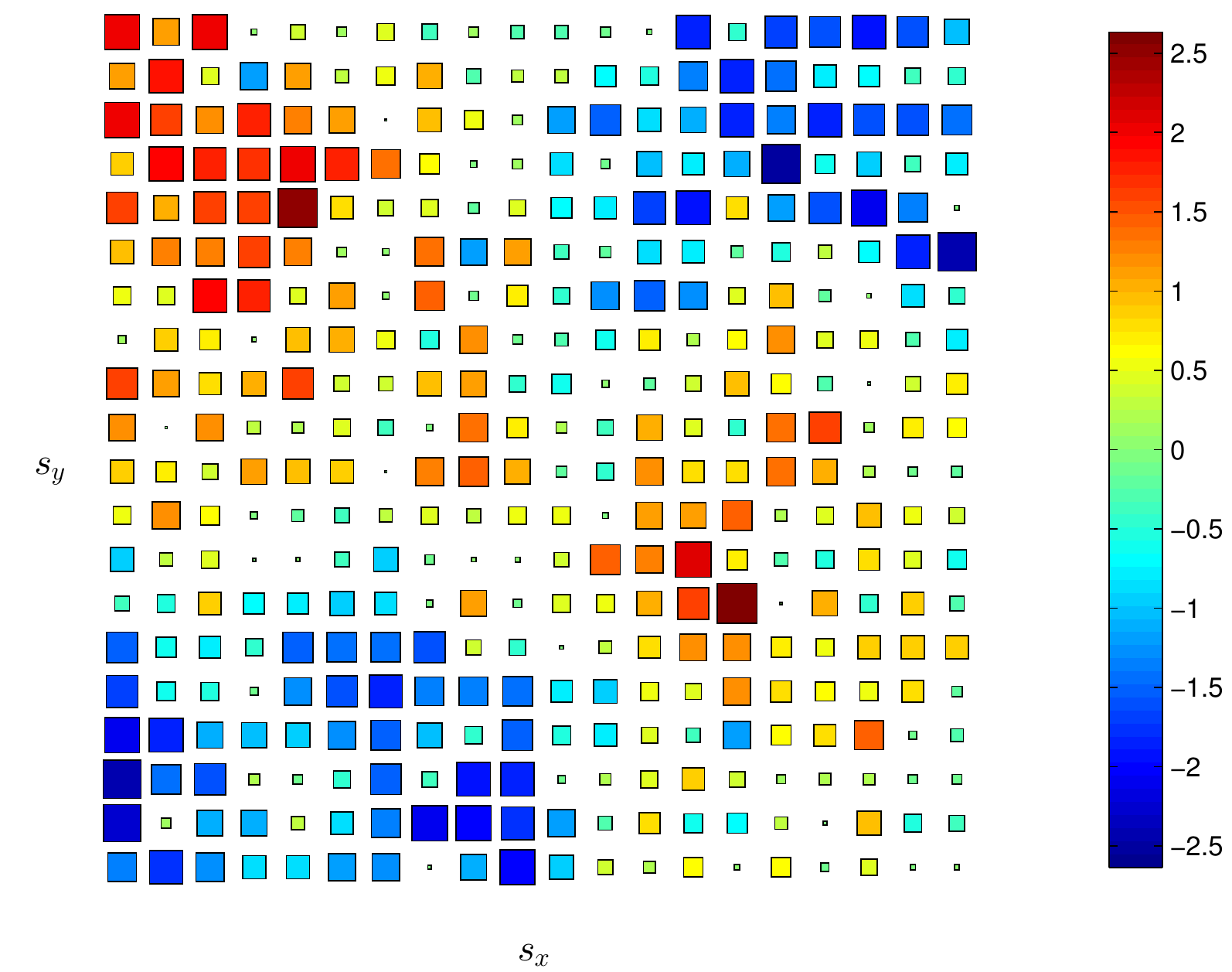}} 
  \subfigure[$y$]{
  \includegraphics[width=0.48\textwidth]{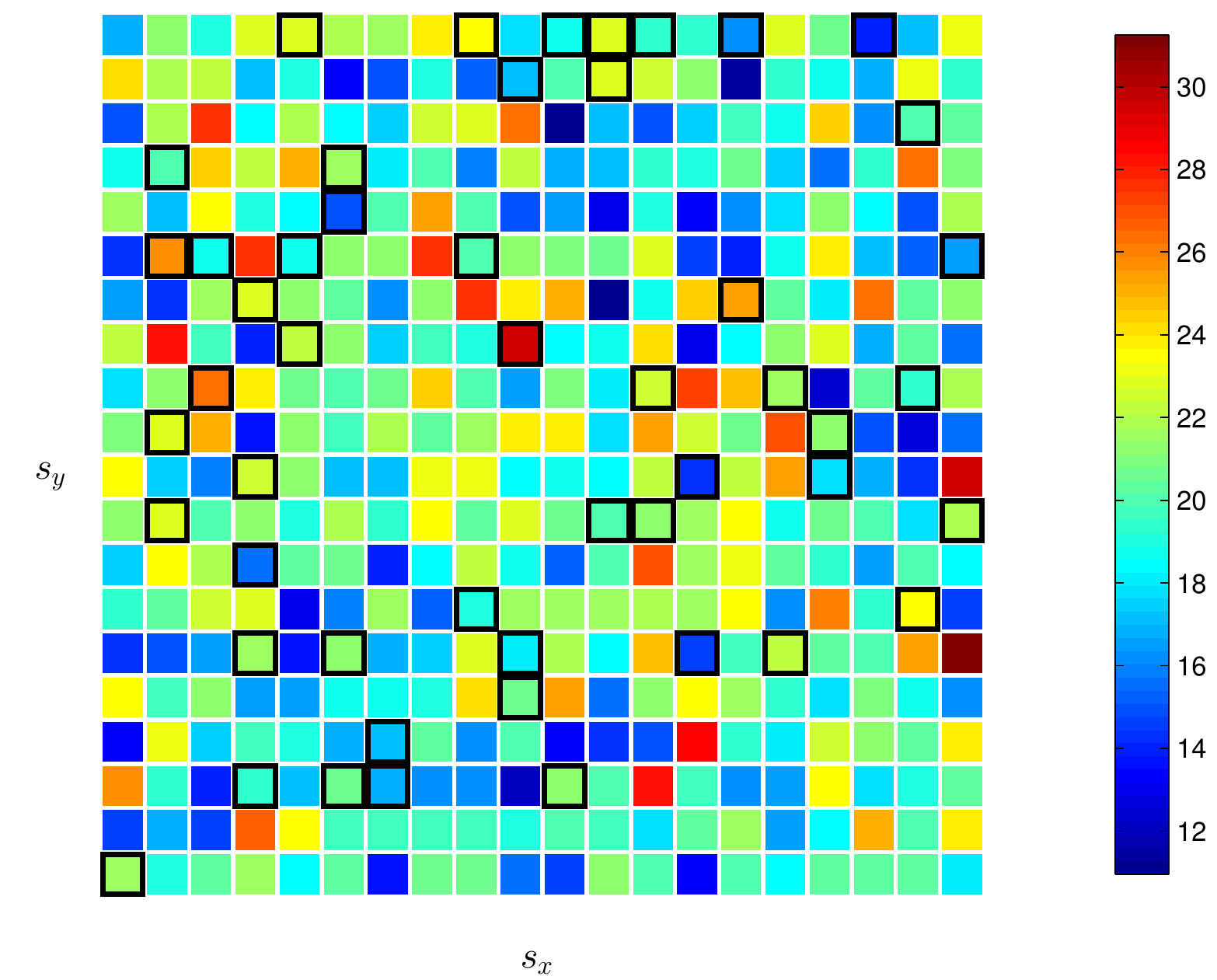}} 
  \caption{Geographically distributed synthetic data: The generated random effects are shown in panel~(a) and the synthetic field measurements (response) in panel~(b). A randomly chosen training set of 50 cells is shown with black squares. Parameters $\tau^2 = 0.5^2, \sigma^2 = 1^2, \phi = 0.5$ were used for the random effects. The width of the area is 1.0 units.\label{fig:synt_map}}
\end{figure}

For synthetic data we generate a set of predictors with only a few of which are important for explaining the variability in the response variable. All the predictor variables are linearly combined so that we need them all in the analysis but there are only a few components that explain the variability. The predictors are located in a 2-dimensional regular grid and spatially correlated random noise is added to the simulated measurements. Further details on how the data are generated are given in \ref{sec:Asimu}. The training set is a randomly selected subset of all the locations and for the rest, response variables are predicted by the model.
Spatially correlated noise is added to the response according to the model given in equation~(\ref{eq:linmodel}) using a structural covariance matrix $\mat{C}$ defined in equation~(\ref{equ:sigmaeta}). 
The spatially correlated residuals are generated with different sets of spatial parameters $\lbrace \tau^2, \sigma^2, \phi\rbrace$ for different degrees of spatial correlation. For the spatial locations of the test area, a $20\times 20$ grid of cells is used (the total number of cells is thus $N=400$) such that the coordinates in  the two directions, $s_x$ and $s_y$, range from $-0.5$ to $0.5$. In each cell $i=1,\dots,N$, a set of predictors and random effects are generated as described above and the response is generated for each cell as
\begin{equation}
  \vec{y}_i = \mat{Z}_i\vec{\beta} + \vec{\eta}_i + \vec{\epsilon}_i.
\end{equation}
See Figure~\ref{fig:synt_map} for an example of the synthetic data.
The principal component predictors, $\mat{X}$, are formed using all the correlated predictors, $\mat{Z}$, i.e., predictors both in the training set and validation set. Only the training set is used in the model parameter estimation.

We test the performance of seven models: traditional Principal Component Regression model using all the orthogonal predictors $\mat{X}$ (equation~(\ref{eq:x})) without assuming spatial effects (labelled as ``PCR'' here) and with spatial effects (``PCRs''), Principal Component Regression model using five first principal components {(five first principal components are used to generate the real response, see \ref{sec:Asimu})} without spatial effects (``PCR6'' -- five principal components plus the intercept) and with spatial effects (``PCR6s''), {the Bayesian Principal Component Regression without spatial effects generated by the original point estimate based regularization model (``BtSVD'' -- see \citet{junttila14}) and its MCMC-method based counterpart (``BPCR'') and the previous MCMC-based regularization model with spatial effects (``BPCRs'').} Models PCR and PCR6 use the ordinary least squares fit. Models PCRs and PCR6s use the MCMC approach with uninformative prior for $\vec{\beta}$. Model BtSVD use the type II - likelihood method to generate point estimate of the model parameters. The models are listed in Table~\ref{tab:models}. 

\begin{table}
  \caption{Different calibration models tested.\label{tab:models}}
 \centering
  \begin{tabular}{l|c|c}
    acronym & no of components & spatial effect \\\hline
    PCR   & all & no  \\
    PCRs  & all & yes \\
    PCR6  & 6   & no \\
    PCR6s  & 6   & yes \\
    BPCR  & all(reqularized) & no \\
    BPCRs & all(reqularized) & yes \\
    BtSVD & all(reqularized) & no
  \end{tabular}
\end{table}

The regression prior hyper parameters used in the experiments are
(see equation~(\ref{eq:alphapri})): $\nu_0 = 1,\ \nu_1 = 0.1$, $a_0 = (\overline{y}/2)^2$, $a_1 = 0.1^2$, where $\overline{y}$ is the mean of the response values in the training set. 
The spatial prior parameters used in the experiments (see equation~(\ref{equ:thetapri})) are
$\mu_\theta = \left[\sigma_\mathrm{PCR}^2/2, \sigma_\mathrm{PCR}^2/2, d_\mathrm{max}/3\right]$ and  $ \sigma_\theta = \left[0.25, 0.25, \infty \right]$, where $\sigma_\mathrm{PCR}^2$ is the residual variance of the BPCR model (without spatial effect), $d_\mathrm{max}$ is the maximum geographical distance in the area, and for $\phi$ there are additional bounds $\left[0.0001\cdot d_\mathrm{max}, d_\mathrm{max}/3 \right]$. 

For validation of performance of the different models, random sets of $n,\ n<N,$ cells are chosen as the model training set, and the other $N-n$ cells serve as the validation set. The model performances are estimated using the predictions in the validation locations. This procedure is repeated 20 times for each set of the spatial parameters with new synthetic data and training set indexes. The statistics of the model performances are collected and displayed in the Results Section.

\subsection{Forest inventory data}\label{test:forest}

\begin{figure}
  \centering
  \includegraphics[width=0.75\textwidth]{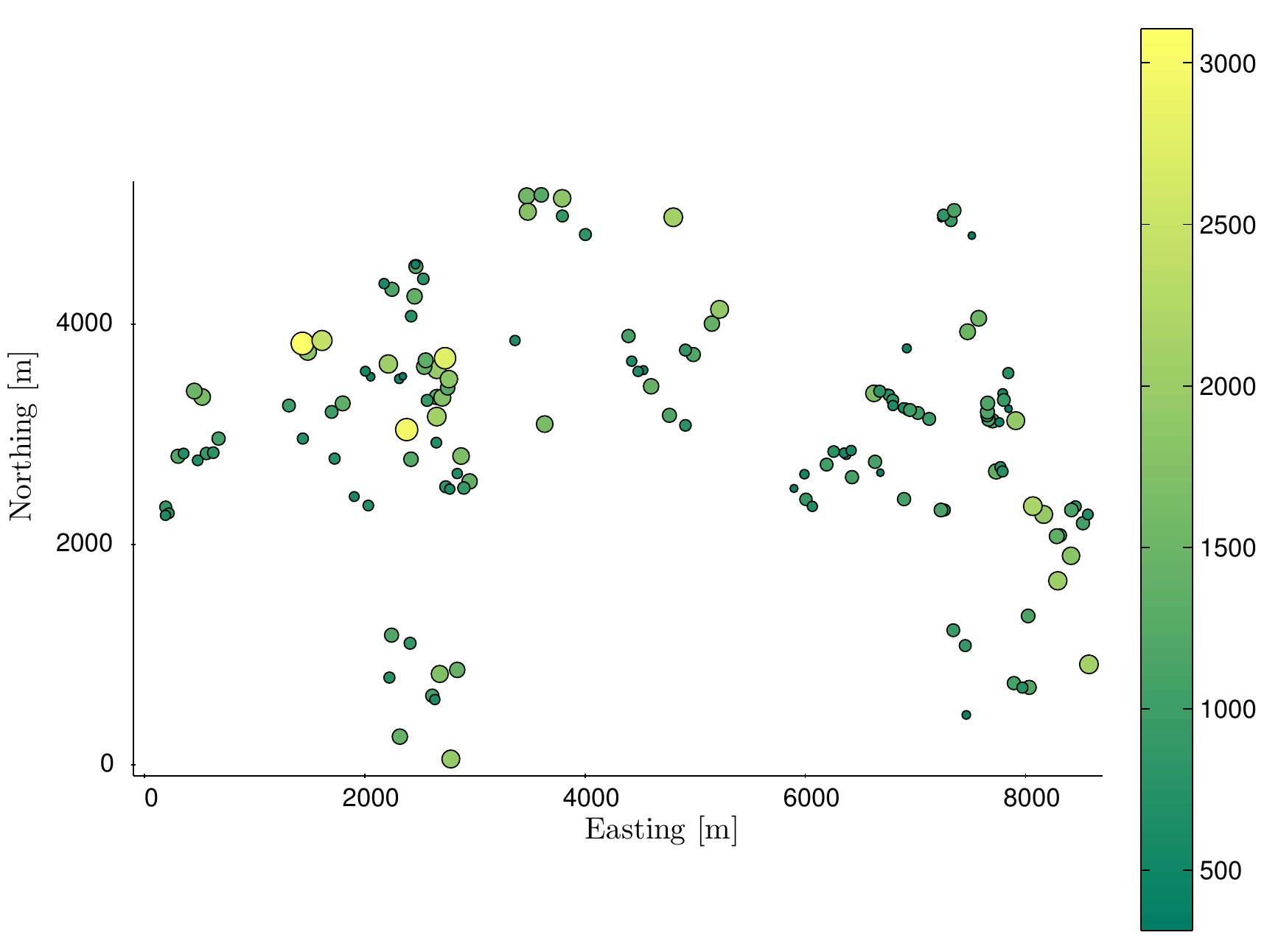}
  \caption{Field plot locations of the forest site Lieksa. The color and size of the circle show the magnitude of the average stem number [number of stems/hectare] of each plot. \label{fig:lieksa_map}}
\end{figure}

Data from a forest site in Lieksa, Eastern Finland is used to validate the performance of the proposed model, see \citet{junttila:2012} for more details. The geographical range of the study area is approximately 8.4 km from east to west, and 5.2 km from south to north. The plot level values of the variable of interest are measured in a set of $N=150$ circular sample plots of 9 m radius {selected according a simple random sampling strategy,}  each and stored in a response vector $\mat{\by}$. The average distance between the field plots is 134.6 m, varying between 17.4 m and 813.1 m. The plot locations given as plot centre coordinates are shown in Figure~\ref{fig:lieksa_map}.

The plot-level LiDAR predictors, $\mat{Z}$, are estimated from airborne LiDAR remote sensing observations. The predictors are extracted from laser scanning point cloud data clipped to the extent of a circular field sample plot with radius 9 m, to match the field measurement data. A total of 38 LiDAR predictors are estimated. The predictors are different statistics of the LiDAR pulse data, such as percentiles of the first and last pulse height and intensity, mean and standard deviation of the first pulse height, and proportion of the ground hits of the first and last pulse heights. The condition number of the resulting predictor matrix was over 22,000 meaning severe multicollinearity. After the orthonormalization procedure described in Section~\ref{sec:pca}, and leaving out those columns with almost zero singular values, we were left with 19 predictors, $\mat{X}$, still explaining practically all of the original predictor variability. 

The LiDAR predictors are known to correlate well with different ecological variables used in forestry. In this study, the variable stem number per hectare is used because it is the most challenging variable of interest here (the worst prediction precision) and spatial correlation in the linear model residuals exists. Figure~\ref{fig:lieksa_map} shows the range and location of the stem number measurements.

To validate the model performance, a Leave-One-Out (LOO) cross-validation procedure is used. Each field plot out of the available $N$ plots serves as the validation plot at a time and the rest of the $N-1$ plots will serve as candidate plots from which a random training set of size $n$, $n\leq N-1$, is selected. The model parameters are estimated using the training set with the in-situ observed stem number, the principal component predictor values and the geographical coordinates of the training set plots. The stem number is predicted for the validation plot using the estimated model with validation plot geographical location and the principal component predictors.

The training set plot locations are selected from the available field plots so that the spatial coverage is as wide as possible. A MaxiMin criterion which maximizes the minimum geographical distance between the $n$ plots is used, and the training set for each iteration is selected using Simulated Annealing method as described in \citet{junttila13}.
A good spatial coverage of the training set plots is needed to detect and use the possible spatial correlation in the prediction in an optimal manner.

We verify six different models. The model performance is validated for the principal component regression with all the 18 predictors without spatial effects (PCR); PCR with spatial effects (PCRs); PCR with Bayesian regularization of the predictors without spatial effects (BPCR); and the proposed model, BPCR combined with the spatial effects (BPCRs). In this study, we also test the prediction performance of PCR with limited number of predictors so that the number of components is estimated using the training set at hand. For each training set, all possible PCR$(q+1)$, $q=1,2,\ldots,18$, models are estimated. The selected model is such that it has the minimum number of principal components required to give root mean square error at most 5\% larger than $\mathrm{RMSE}_\mathrm{min}$, the minimum for all models. This model is labelled as ``PCR adapt''. The corresponding model where the spatial effects are included, is labelled as ``PCRs adapt''.

\subsection{Statistics of model performance}

The prediction precision and accuracy of each training set size are estimated with relative root mean square error, 
\begin{equation}
  \label{equ:rmse}
  \mathrm{RMSE\%} = \frac{\sqrt{\sum_{i=1}^{N}(\widehat{y}_i-y_i)^2/N}}{\overline{\vec{y}}} \times 100
\end{equation}
and relative bias,
\begin{equation}
  \label{equ:bias}
  \mathrm{Bias\%} = \frac{\sum_{i=1}^{N}(\widehat{y}_i-y_i)/N}{\overline{\vec{y}}}  \times 100
\end{equation}
where $\widehat{y}_i$ is the predicted value for location $i$ and $\overline{\vec{y}}$ is the mean over all measured values ${y}_i$. We also estimate the $Q^2$ value for cross-validated coefficient of determination:
\begin{equation}
 Q^2 = \left(1- \frac{\sum_{i=1}^N(\widehat{y}_i-y_i)^2}{\sum_{i=1}^N(y_i-\overline{\vec{y}})^2}\right)\times 100.
\end{equation} 
It is similar to the ordinary coefficient of determination, $R^2$. The value is 100(\%), if the model can predict the validation set perfectly. The $Q^2$ can become even negative if the model performs worse in prediction than the most simple model, mean of the training set measurements.

\section{Results}
\label{sec:results}

\subsection{Synthetic data}

Model performances are tested against different types of synthetic data with varying degrees of spatial correlation in random residuals of data, $\phi = \lbrace 0.001, 0.1, 0.5\rbrace$ and sizes of random training set, $n = \lbrace 20,30,50,80,120\rbrace$.  Each dataset is used for the model calibration with all the different methods and the predictions are compared to the synthetic truth. For each combination, 20 repetitions are made to collect average behaviour.

\begin{figure}
  \centering
  \subfigure[PCR]{
  \includegraphics[width=0.45\textwidth]{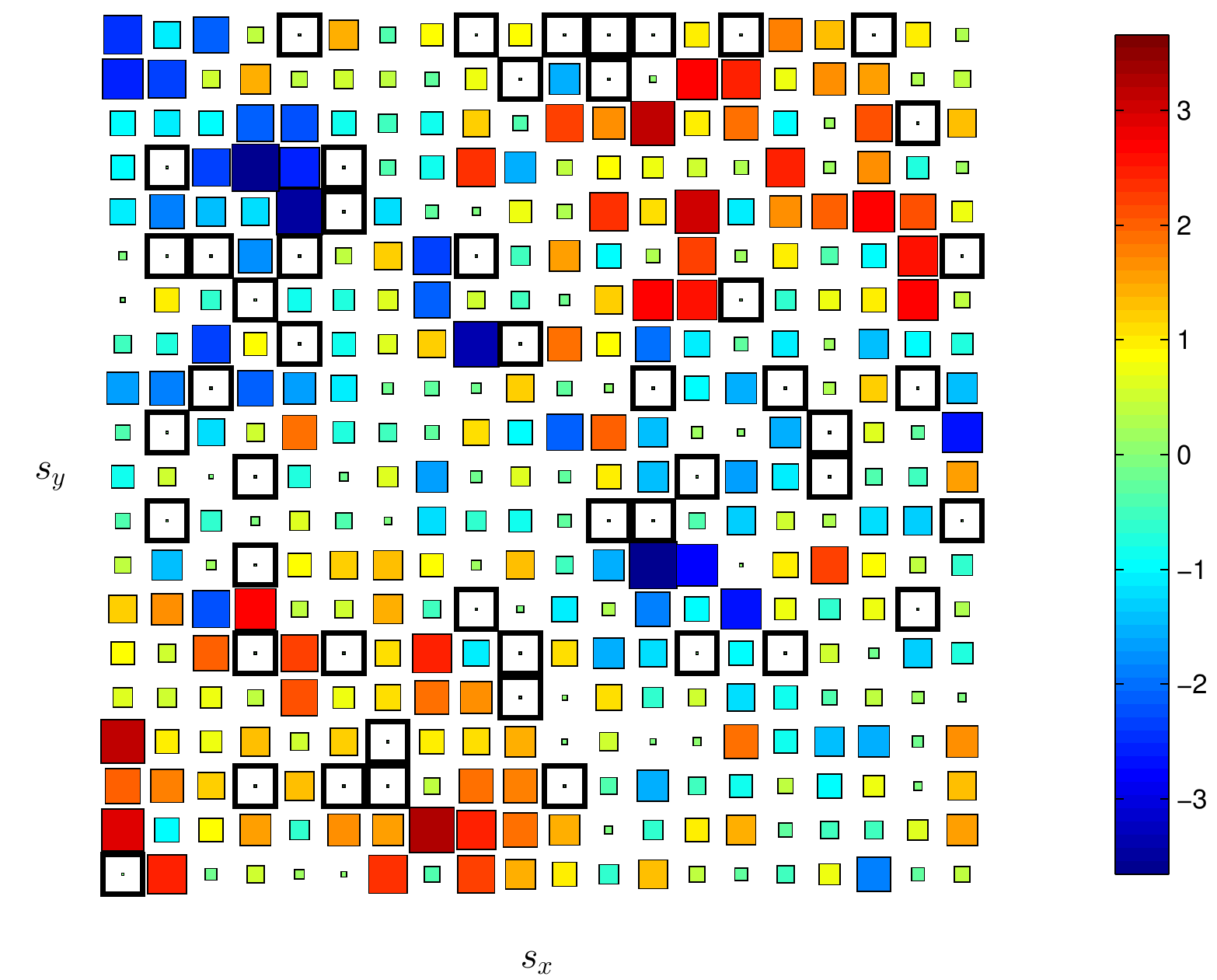}\label{fig:pcr}}
  \subfigure[PCRs]{
  \includegraphics[width=0.45\textwidth]{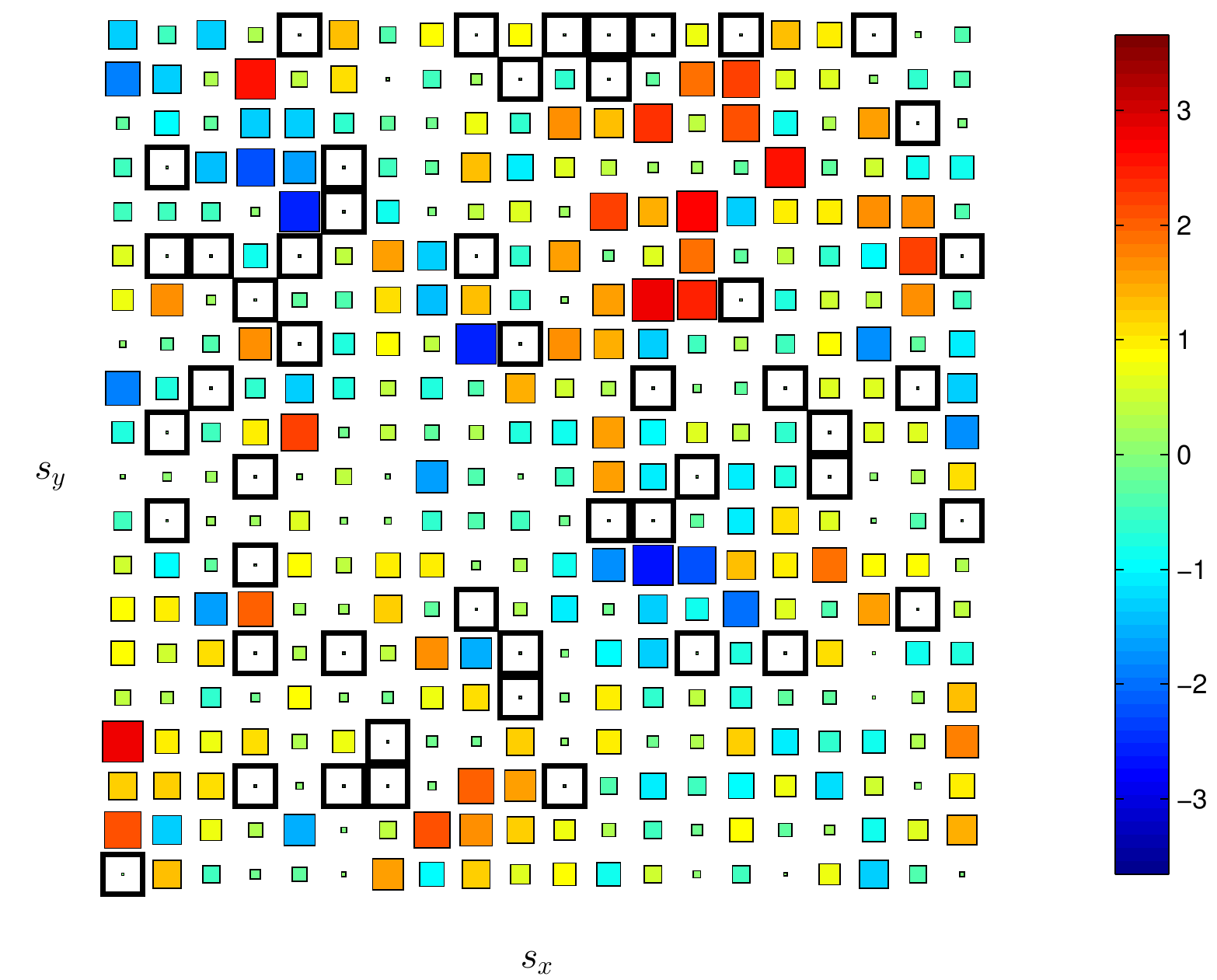}\label{fig:pcrs}} 
  \subfigure[PCR6]{
  \includegraphics[width=0.45\textwidth]{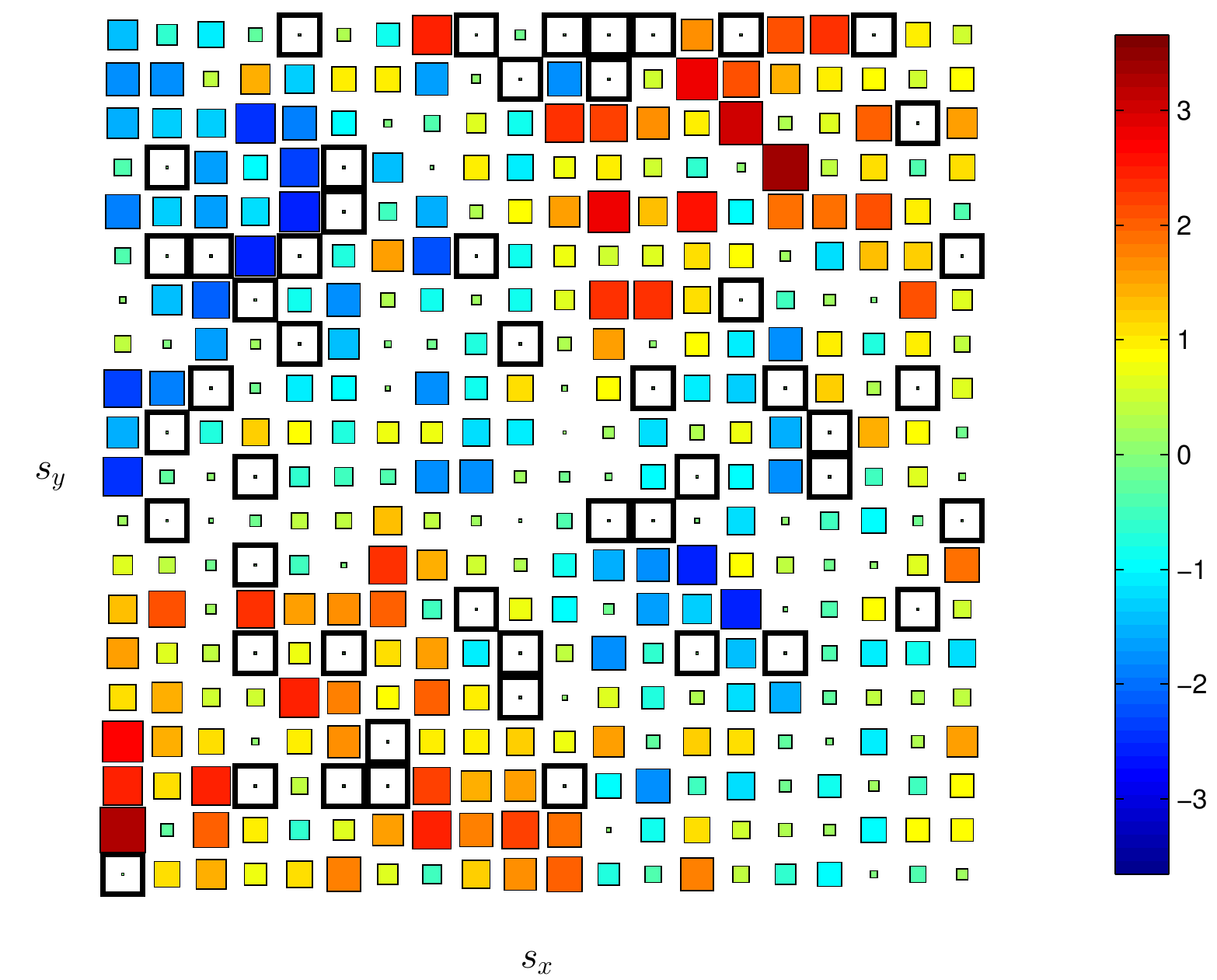}\label{fig:pcr6}} 
  \subfigure[PCR6s]{
  \includegraphics[width=0.45\textwidth]{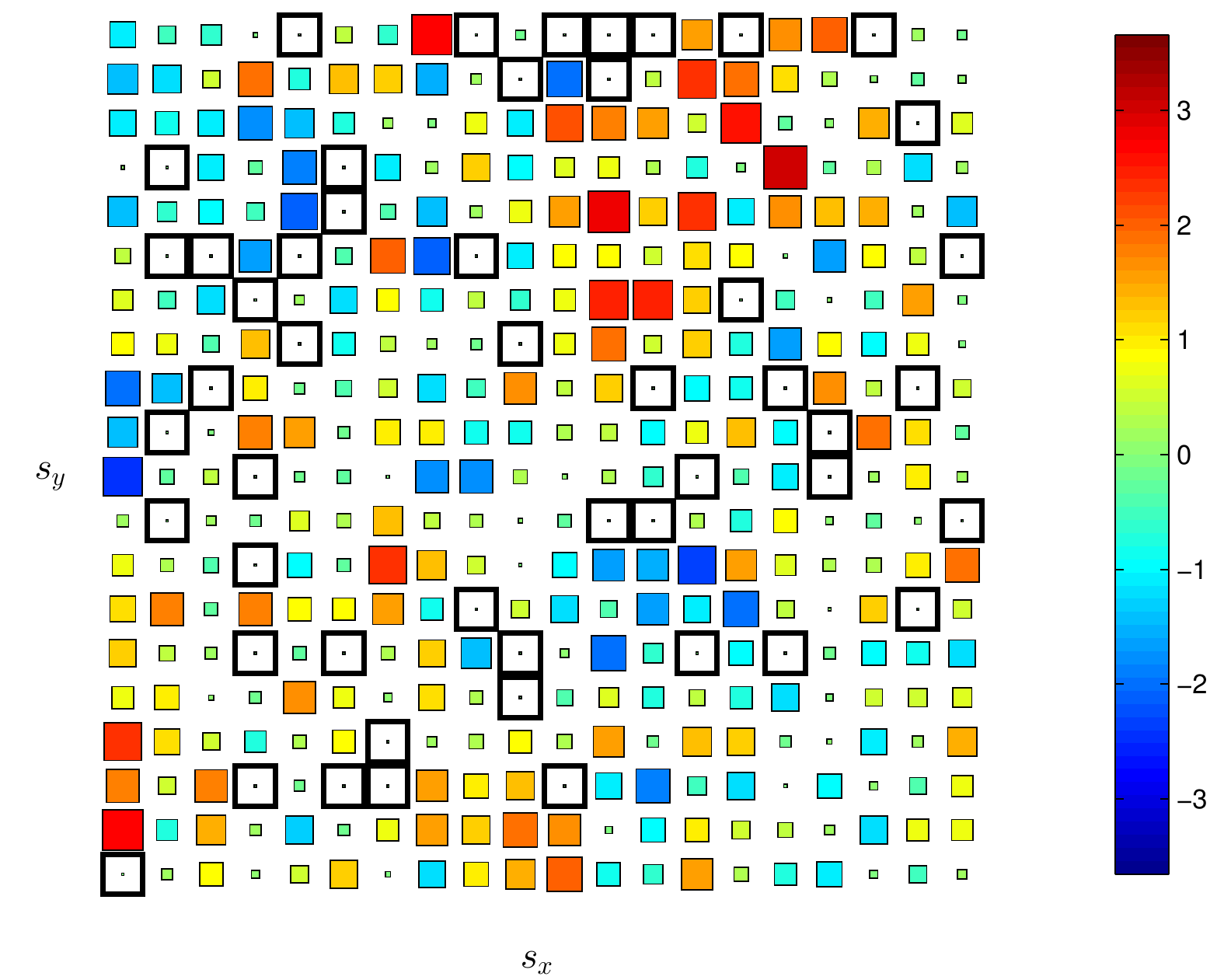}\label{fig:pcr6s}} 
  \subfigure[BPCR]{
  \includegraphics[width=0.45\textwidth]{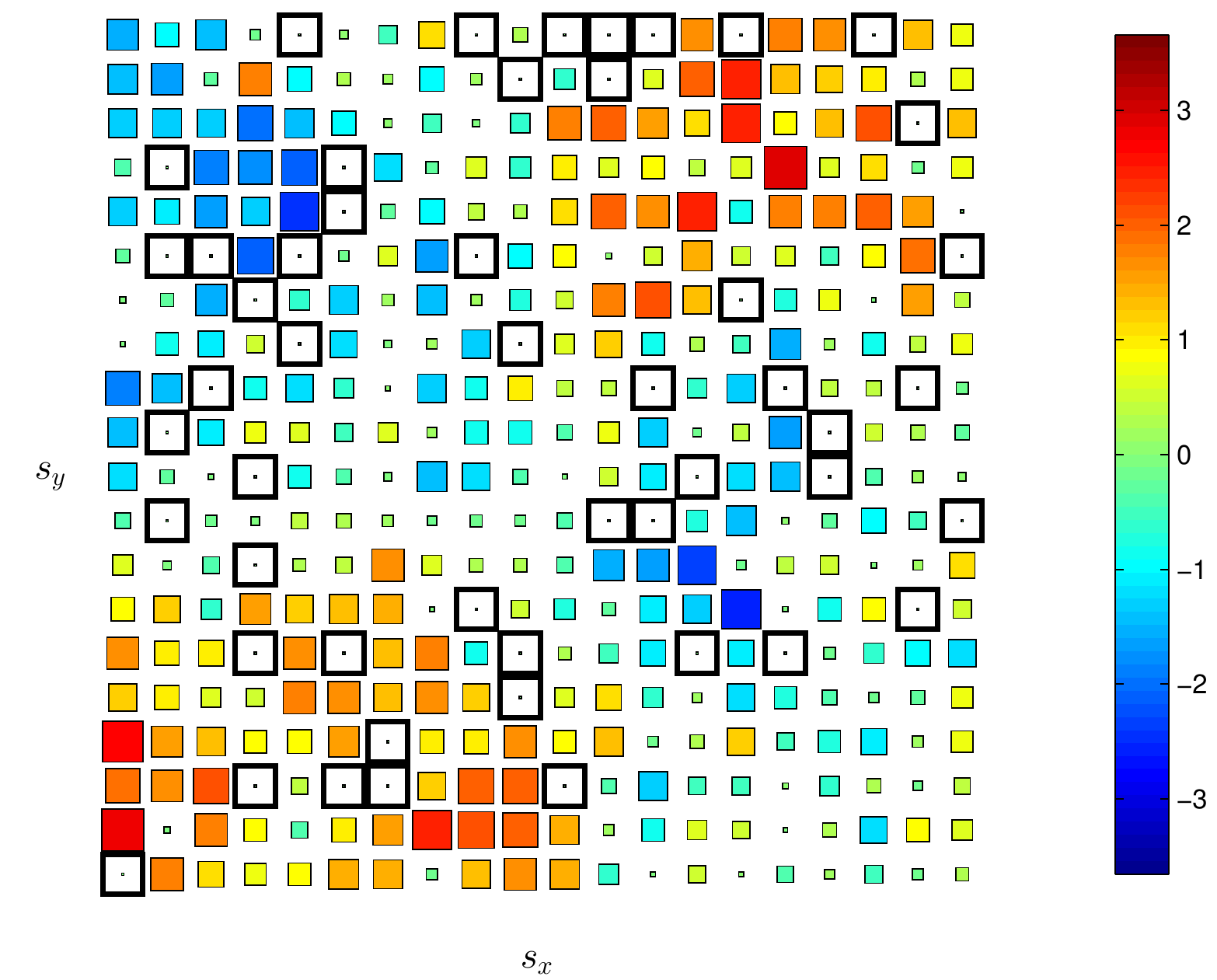}\label{fig:bpcr}} 
  \subfigure[BPCRs]{
  \includegraphics[width=0.45\textwidth]{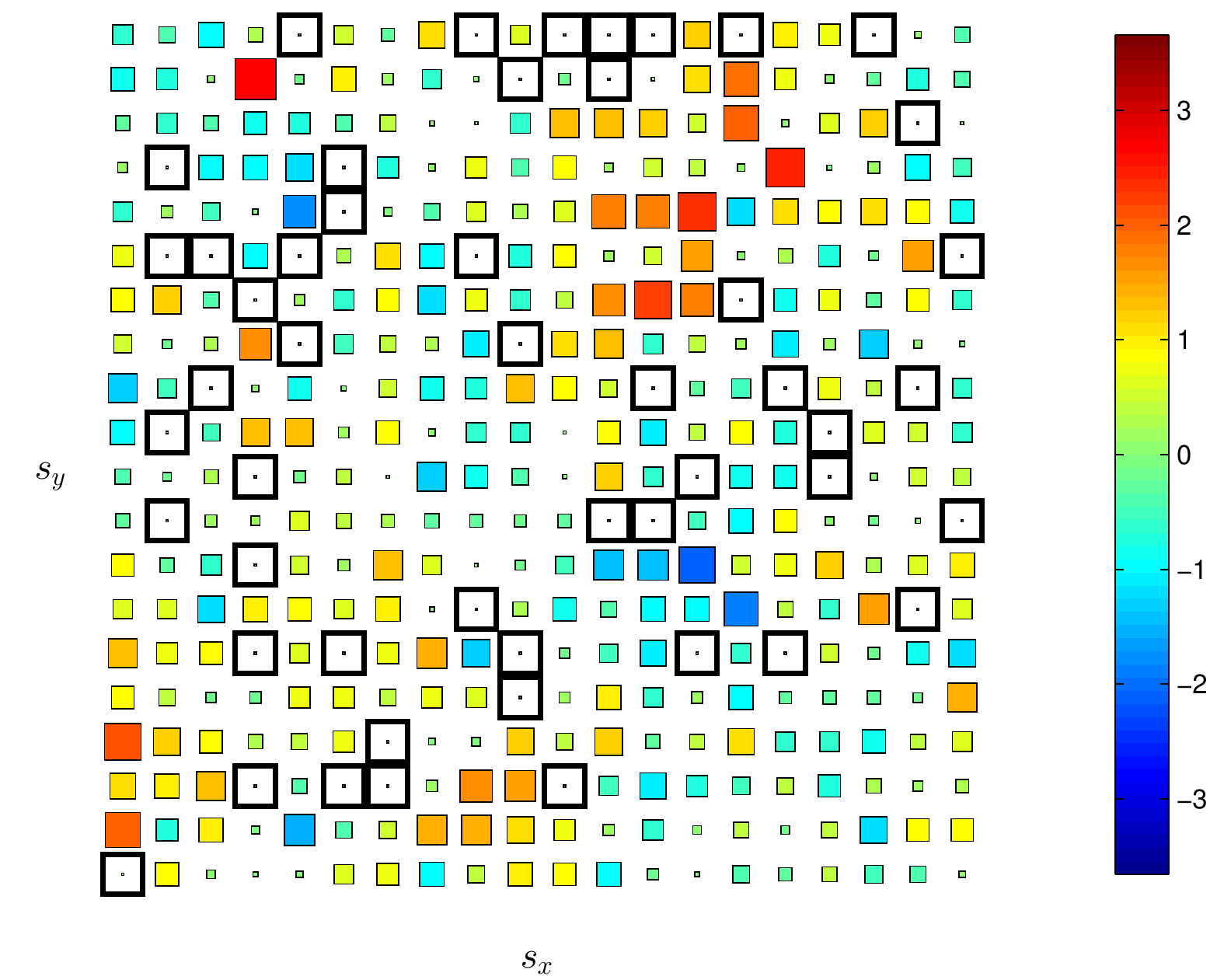}\label{fig:bpcrs}} 
  \caption{Prediction residuals of different models using the synthetic data shown in Figure~\ref{fig:synt_map}. The cell sizes in the figures are relative to the absolute residual value in the cell, the color defines also the sign of the prediction error. The training set cells used for model parameter estimation ($n=50$) are shown with black squares. The residual maps on the left column (subfigures a, c and e) are generated with models without spatial effect, PCR, PCR6 and BPCR. On the right column (subfigures b, d and f), the maps of corresponding models with spatial effects are shown.  \label{fig:synt_map_res}}
\end{figure}

To visualise the results with one synthetic test dataset, the dataset shown already in Figure~\ref{fig:synt_map} is used. This dataset is generated with parameter values $\tau = 0.5,\ \sigma = 1$ and $\phi = 0.5$, thus there is large spatial correlation. The size of training set is 50 cells. The map of prediction residuals and the error statistics for different models are shown in Figure~\ref{fig:synt_map_res} and Table~\ref{tab:synt_example}. 

Among the models without spatial effects, PCR, PCR6 and BPCR, the prediction residuals of the model PCR have the most variability, more than the residuals of the models PCR6 and BPCR, see Figures~\ref{fig:pcr}, \ref{fig:pcr6} and \ref{fig:bpcr}. With all these models, the residual surface map shows spatial patterns: the cell level residuals are approximately the opposite of those in actual spatial effect map in Figure~\ref{fig:synt_map}.a). The magnitude of these residuals is closest to zero with model BPCR, which can also be seen in Table~\ref{tab:synt_example} where the RMSE\% and $Q^2$ of BPCR are 5.55\% and 90.35, correspondingly, while for PCR they are 6.76\% and 85.70 and for PCR6 6.32\% and 87.50. 

The residuals of the models with spatial effect included, PCRs, PCR6s and BPCRs, deviate less from zero than the residuals of corresponding models without spatial effects, see Figures~\ref{fig:pcrs}, \ref{fig:pcr6s} and \ref{fig:bpcrs}. For BPCRs, the residual surface is closest to zero and its geographical structure does no longer resemble that in Figure~\ref{fig:synt_map}.a). It outperforms the models PCRs and PCR6s, as seen in Table~\ref{tab:synt_example}. The RMSE\% and $Q^2$ of BPCRs are 4.23\% and 94.40, correspondingly, while for PCRs they are 5.20\% and 91.55 and PCR6s 5.46\% and 90.68. In fact, with this training set size, model PCRs gives better results than the truncated model PCR6s.

\begin{table}
  \caption{Error statistics of the synthetic data example shown in Figures~\ref{fig:synt_map} and~\ref{fig:synt_map_res}.\label{tab:synt_example}}
  \centering
    \begin{tabular}{l c c c c c c}
      & PCR & PCRs & PCR6 & PCR6s & BPCR & BPCRs \\\hline
      Bias\% & 0.68 & 0.68 & 0.95 & 0.86 & 0.84 & 0.76 \\
      RMSE\% & 6.76 & 5.20 & 6.32 & 5.46 & 5.55 & 4.23 \\
      $Q^2$ & 85.70 & 91.55 & 87.50 & 90.68 & 90.35 & 94.40 \\
    \end{tabular}
\end{table}

Error statistics in Table~\ref{tab:synt_example} show that Bias\% is close to zero for each model. This holds true also in the later results of synthetic data and thus it is not shown in the results separately. 

\begin{figure}
  \centering
  \subfigure[$\tau^2 = 0.25, \sigma^2 = 1, \phi = 0.001 $]{
  \includegraphics[width=0.48\textwidth]{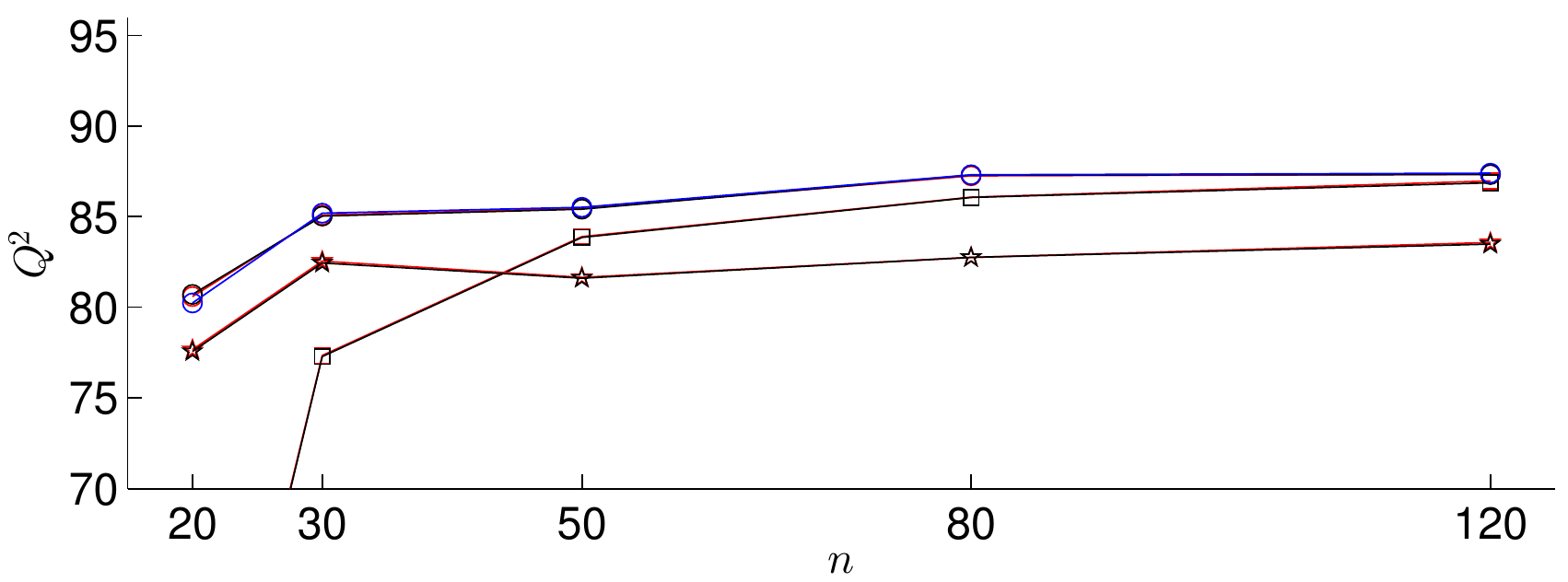}} 
  \subfigure[$\tau^2 = 0.25, \sigma^2 = 1, \phi = 0.001 $]{
  \includegraphics[width=0.48\textwidth]{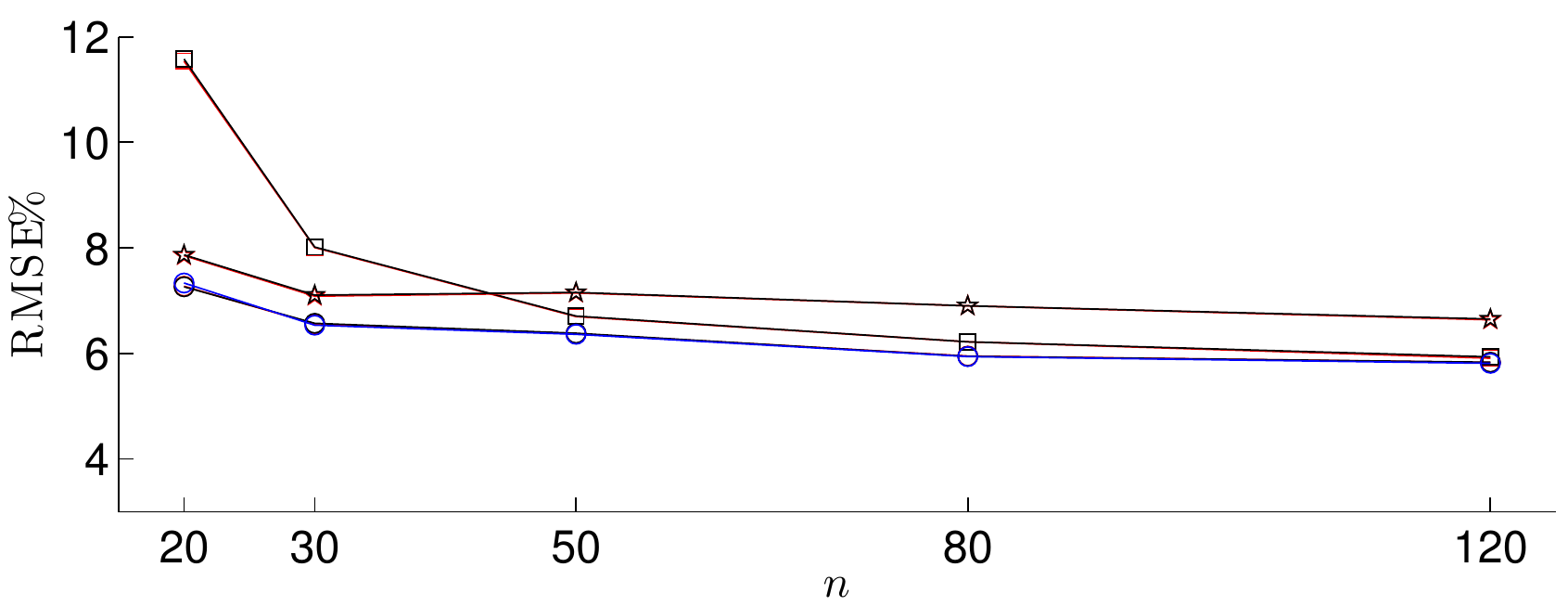}} 
  \subfigure[$\tau^2 = 0.25, \sigma^2 = 1, \phi = 0.1 $]{
  \includegraphics[width=0.48\textwidth]{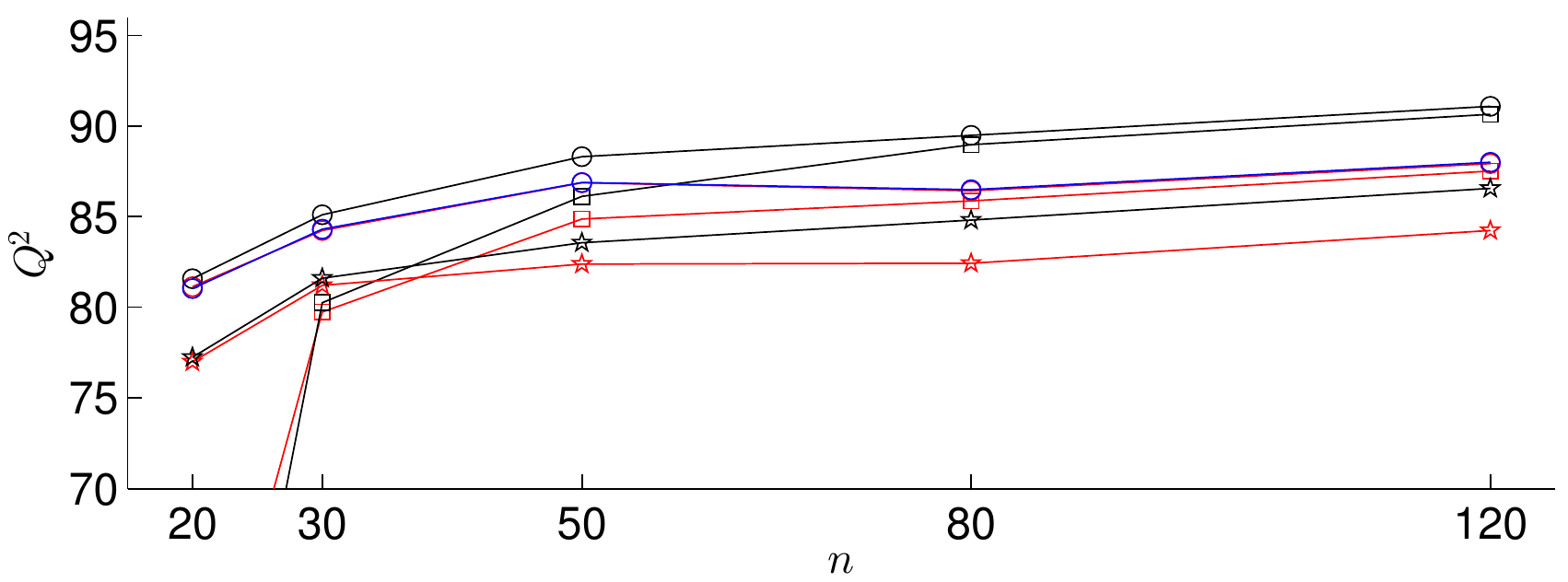}} 
  \subfigure[$\tau^2 = 0.25, \sigma^2 = 1, \phi = 0.1 $]{
  \includegraphics[width=0.48\textwidth]{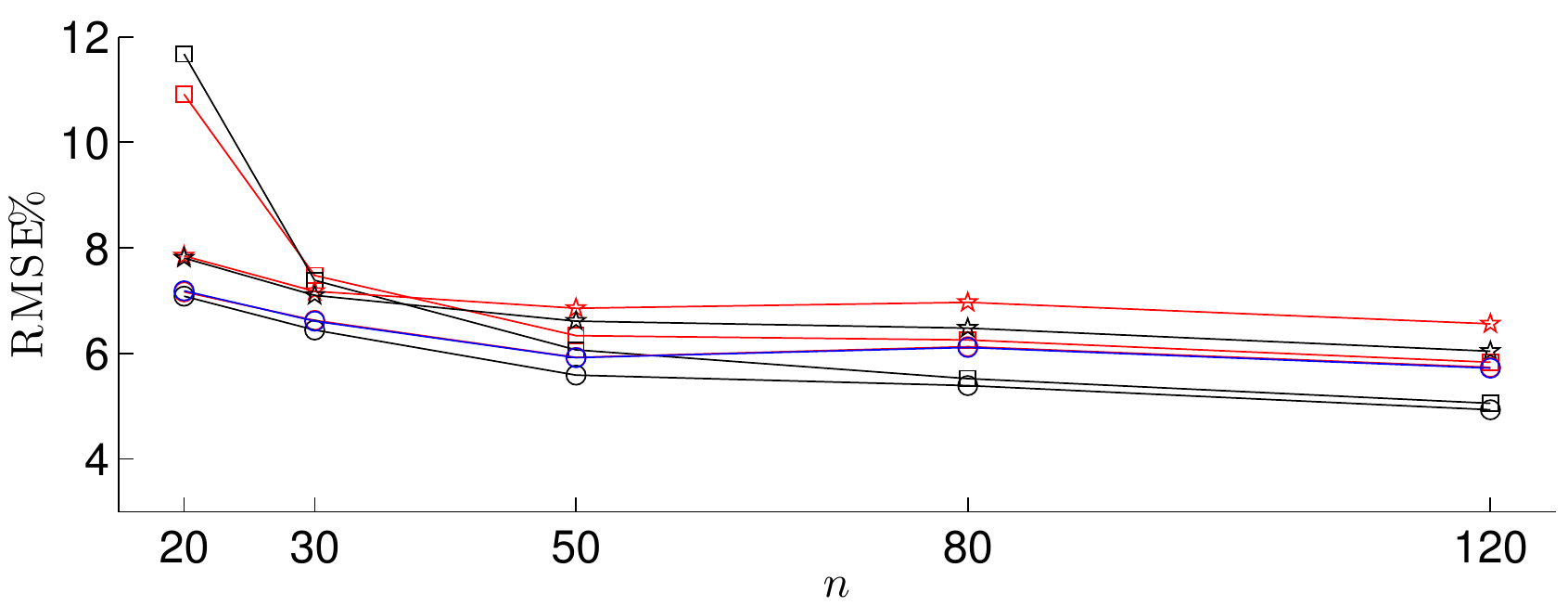}} 
  \subfigure[$\tau^2 = 0.25, \sigma^2 = 1, \phi = 0.5 $]{
  \includegraphics[width=0.48\textwidth]{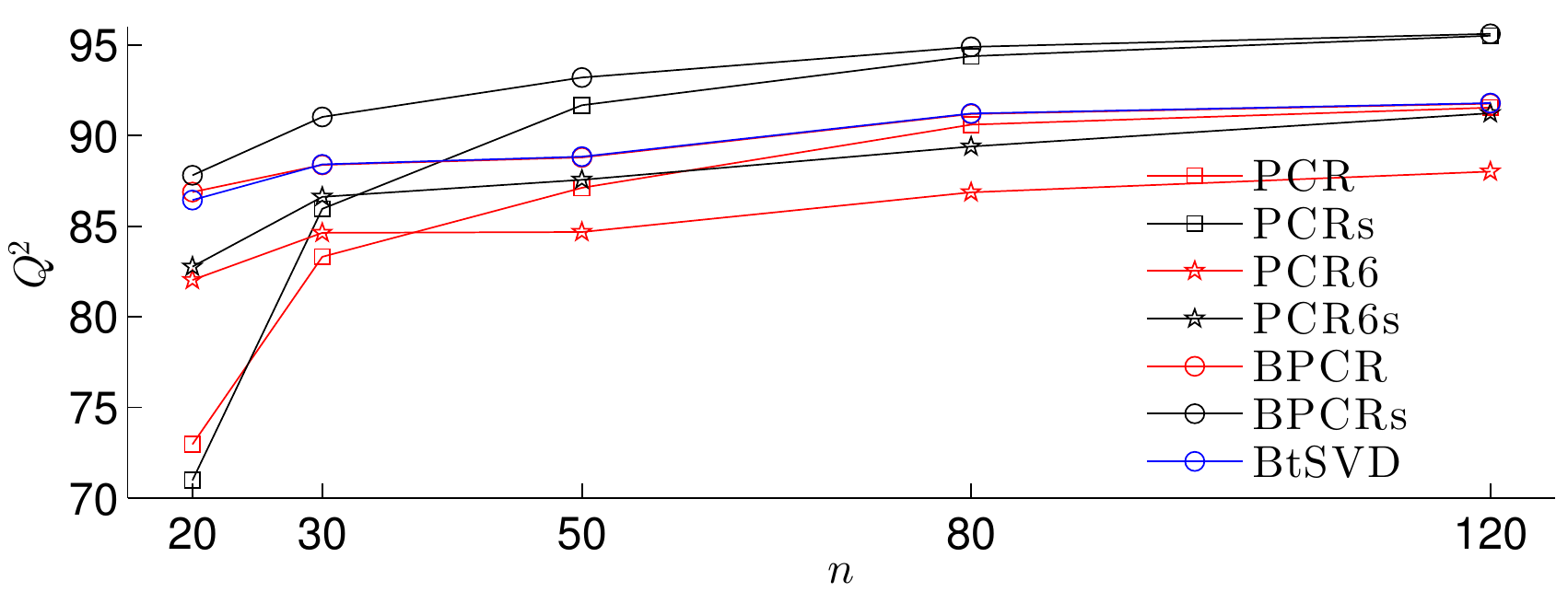}} 
  \subfigure[$\tau^2 = 0.25, \sigma^2 = 1, \phi = 0.5 $]{
  \includegraphics[width=0.48\textwidth]{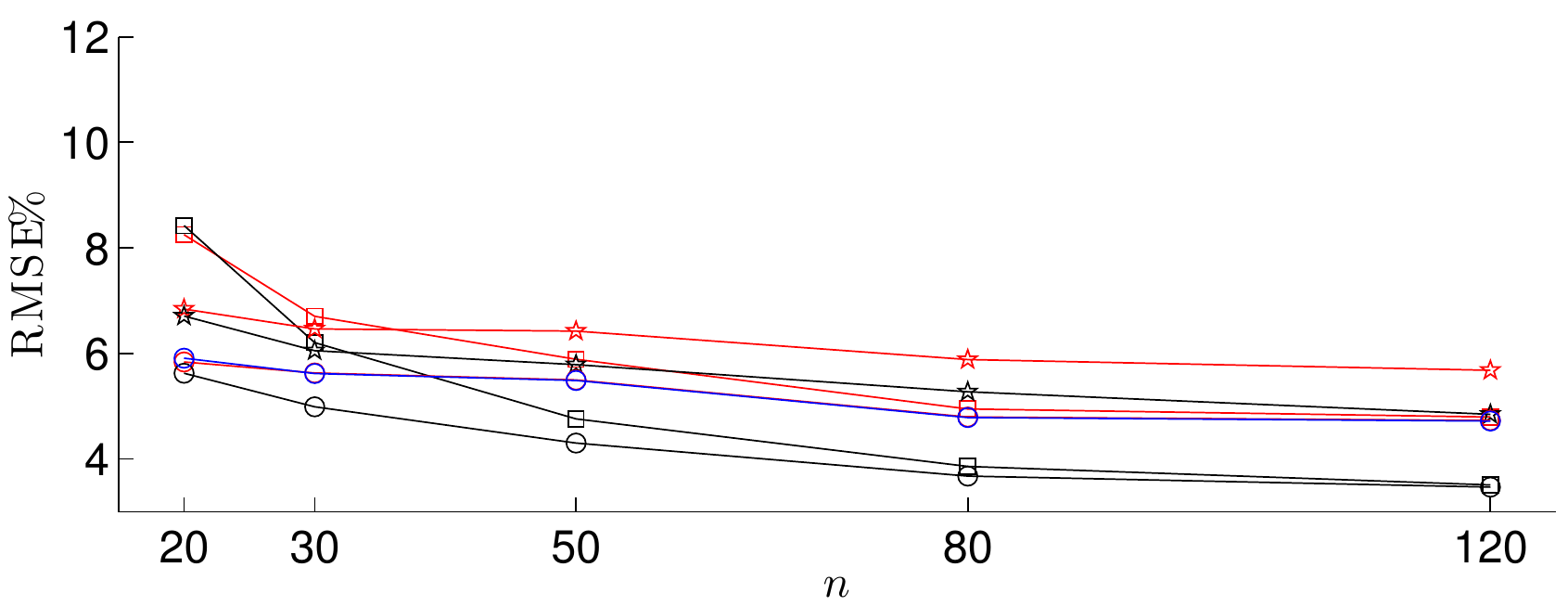}} 
  \caption{Average $Q^2$ and RMSE\% of the predictions of the synthetic data generated with different parameters.  \label{fig:synt_map_res_all}}
\end{figure}

The overall performance of the models with different training set sizes are shown in Figure~\ref{fig:synt_map_res_all}.  

Among the models without spatial effect the model PCR gives a quite good RMSE\% when the training set size is large ($n=120$). However, with smaller sizes, the predictions deteriorate. For $n=20$, the $Q^2$ estimate for PCR is close to zero. This effect of over-training is avoided by limiting the use of predictors, as in models PCR6 and BPCR. Even though the synthetic response data were generated using the same number of principal components, the model PCR6 yields less precise predictions when the training set size is large. This model, however, is useful when the training set size is small. With all the training set data types used in this study and all the training set sizes, PCR6 is always outperformed by model BPCR.
{For all the cases, the model performance of the original Bayesian regularization model, BtSVD, see \cite{junttila14}, is similar to that of its MCMC-method based counterpart, model BPCR. }

In practice, when there is no spatial effect in the synthetic data ($\phi = 0.001$), the corresponding models with and without spatial effect give similar results, see pairs PCR \& PCRs, PCR6 \& PCR6s and BPCR \& BPCRs in Figure~\ref{fig:synt_map_res_all}.a).
However, when the spatial effects in the data become larger, see subfigures b) and c), differences in model performances appear. If the training set size is small and the cell locations are far from each other relative to the range of the spatial correlation, the spatial effect cannot be estimated from the training set. If the effective range of the spatial effect is large, even a small number of training set cells can be used to see this effect. For example, with $\phi = 0.1$, the predictions of model BPCRs are better than those of BPCR with training set sizes $>50$, and with $\phi = 0.5$, the same phenomenon can be seen already with training set sizes $>20$.  Also among these models, the model BPCRs outperforms the other models.

\begin{table}
\caption{Average 95\% confidence interval length / the average coverage percentage of models BtSVD, BPCR and BPCRs with different training set sizes and correlation decay parameters.\label{table:CIs}}
\begin{tabular}{l c |c |c |c}
$\phi$ & $n$ & BtSVD & BPCR  & BPCRs   \\\hline
$0.001 $&20 & 5.20 / 90.04 & 5.63 / 93.86 & 5.59 / 94.32 \\
&30 & 5.28 / 94.90 & 5.18 / 95.78 & 5.14 / 95.70 \\
&50 & 4.84 / 94.30 & 4.37 / 94.54 & 4.37 / 94.49 \\
&80 & 4.81 / 95.74 & 3.92 / 95.66 & 3.93 / 95.70 \\
&120 & 4.50 / 94.99 & 3.19 / 94.48 & 3.19 / 94.38 \\
\hline
$0.1 $ &20 & 4.83 / 89.19 & 5.22 / 92.28 & 5.25 / 93.30 \\
&30 & 4.99 / 93.40 & 4.91 / 94.42 & 4.79 / 94.68 \\
&50 & 4.86 / 95.35 & 4.39 / 95.49 & 4.12 / 95.34 \\
&80 & 4.81 / 95.16 & 3.92 / 94.95 & 3.43 / 94.64 \\
&120 & 4.41 / 95.16 & 3.12 / 94.30 & 2.74 / 94.64 \\
\hline
$0.5 $ &20 & 4.23 / 90.81 & 4.59 / 94.84 & 4.72 / 96.51 \\
&30 & 4.36 / 93.79 & 4.30 / 94.84 & 3.94 / 95.57 \\
&50 & 4.04 / 93.73 & 3.66 / 93.66 & 3.03 / 94.99 \\
&80 & 3.71 / 94.89 & 3.02 / 94.69 & 2.45 / 95.67 \\
&120 & 3.67 / 95.40 & 2.60 / 94.73 & 1.92 / 94.96 \\
\hline
\end{tabular}
\end{table}

To assess prediction accuracy and precision we calculated average 95\% confidence interval lengths and the corresponding average coverage percentages for each prediction generated by the MCMC-based models BPCR and BPCRs and for the point estimate based regularization model BtSVD. The results are shown in Table~\ref{table:CIs}. In general, the confidence intervals cover well the true values of the variable. The percentage is close to 95\% for all the training set sizes and models except for the smallest size, $n=20$, where the point estimate coverage percentage drops to approximately 90\% for some methods. The confidence interval lengths are typically largest for the point estimate based model BtSVD compared to the MCMC-based models. The models BtSVD and BPCR are basically the same, but solved with different methods, and their prediction performances are similar, see Figure~\ref{fig:synt_map_res_all}. However, the average confidence interval lengths of model BPCR are clearly smaller than those of BtSVD. Without spatial correlation in the data, the confidence interval lengths for BPCR and BPCRs are equal, and with spatial correlation, the confidence interval for BPCRs is smaller, which is a reasonable result because the model performance is also better in case of spatial correlation. We can conclude that full Bayesian analysis with MCMC-based uncertainty calculations for the predictions results overall the most accurate and precise results.

\subsection{Forest inventory, stem number prediction}\label{sec:forinv_stemno}

In the forest inventory case, we tested training set sizes of $n=\lbrace 15, 20, 30,\ldots, 60, 80, \ldots, 140, 149\rbrace$.  
With 30 field plots, the average minimum distance among the different random sample plots varies between 550 - 654 m (average 609 m), while with 80 field plots it is 242 - 274 m (average 256 m), and with 140 field plots only 144 -152 m (average 149 m). 

\begin{table}
\caption{Error statistics of the Leave-One-Out models ($n=149$) in the forest inventory case.\label{tab:loo_sols}}
\centering
\begin{tabular}{l| c c| c c| c c}
 & PCR	 & PCRs	 & PCR adapt	 & PCRs adapt	 & BPCR	 & BPCRs	 \\\hline
 Bias\% & 0.61 	 & 0.75 	 & 0.41 	 & 0.11 	 & 0.20 	 & 0.04 	 \\
 RMSE\% & 37.70 	 & 33.96 	 & 35.44 	 & 32.08 	 & 35.61 	 & 32.35 	 \\
 $Q^2$ & 0.45 	 & 0.55 	 & 0.51 	 & 0.60 	 & 0.51 	 & 0.59 	 
\end{tabular}
\end{table}
Prediction performance of the six different models, PCR, PCRs, PCR adapt, PCRs adapt, BPCR and BPCRs, are evaluated with the given data. Table~\ref{tab:loo_sols} shows the baseline results for Bias\%, RMSE\% and $Q^2$ obtained with leave-one-out procedure using all the available 149 field plots as the training set to predict each validation plot. These results show that, with a large enough training set, all models with spatial random effects perform similarly, as well as those without. However, the results of models PCR and PCRs are worse than those estimated with some control, i.e.\ truncation or regularization, on predictors. The best performance is given by the adaptive PCR models, PCR adapt and PCRs adapt, but the difference in performance of the corresponding Bayesian models, BPCR and BPCRs, is very small. The results show that there is spatial correlation in the linear model residuals, and the models including spatial effects give clearly better predictions than the models without spatial effects.

\begin{figure}
  \centering
  \subfigure[Models PCR and PCRs.]{
  \includegraphics[width=0.48\textwidth]{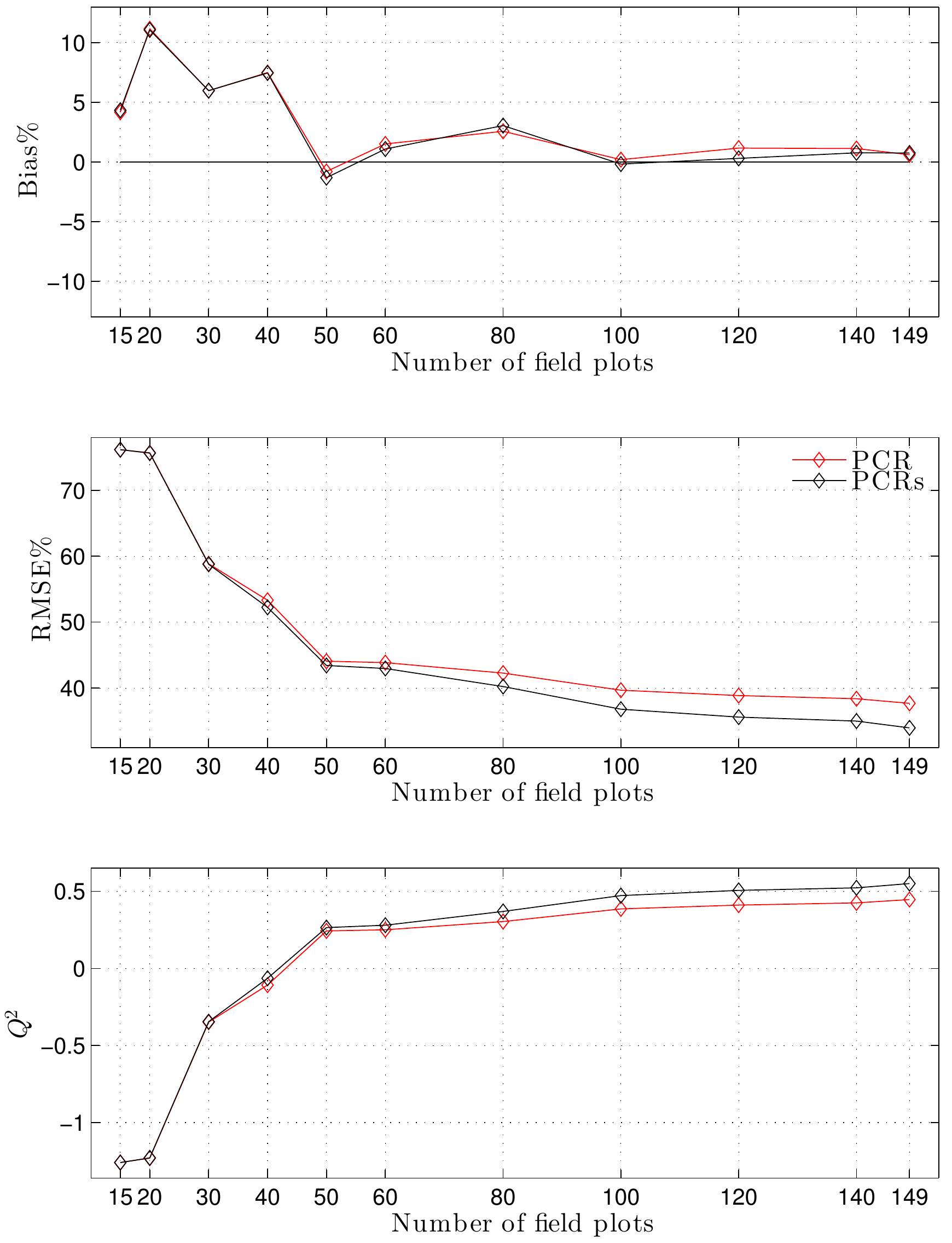}} 
  \subfigure[Models PCR(s) adapt and BPCR(s).]{
  \includegraphics[width=0.48\textwidth]{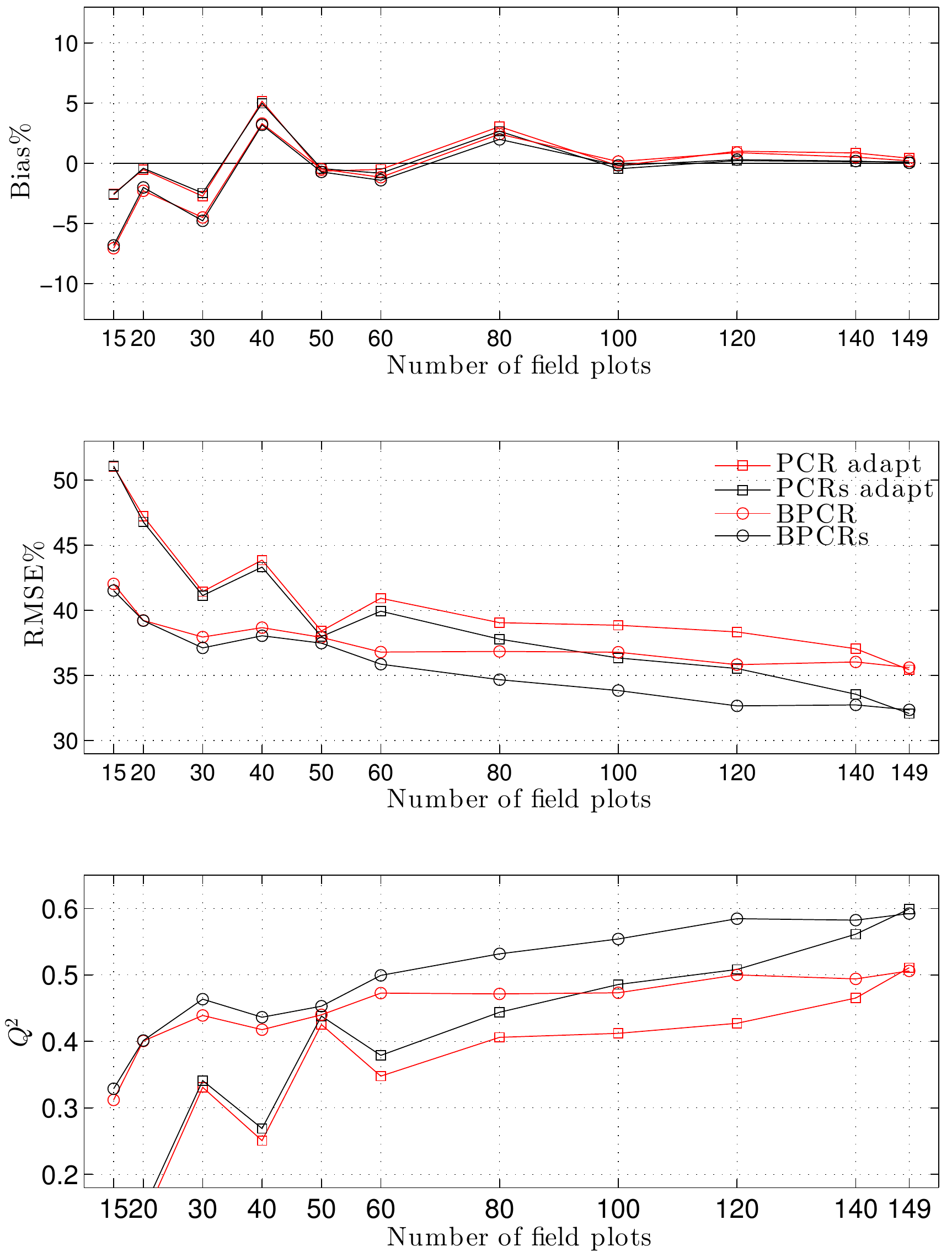}} 
  \caption{Bias\%, RMSE\% and $Q^2$ for different models in the forest inventory case. The left panels (a) show results of models which don't control the use of predictors, and the right panels (b) show results of models that truncate or regularize the use of predictors. Note the different scale on the axes of RMSE\% and $Q^2$ results in left and right panels.  \label{fig:lieksa_res}}
\end{figure}
The corresponding results estimated with different training set sizes are shown in Figure~\ref{fig:lieksa_res}.
The results obtained with PCR and PCRs (the left panels of the Figure~\ref{fig:lieksa_res}) lack the ability to  achieve precise predictions even with a relatively large number of field measurements, and with small training set sizes ($n<40$) the $Q^2$ values become even negative, i.e.\ the models perform worse than the no-model estimate. The results given on the right panels, obtained by controlling the use of predictors (models PCR adapt, PCRs adapt, BPCR, BPCRs) do not suffer from as severe over-training. 
Among these results, the Bayesian models perform better than the adaptive PCR models when $n<149$. 

For the model PCR adapt, the number of principal components used is arbitrarily restricted according to the training set data. Even though the number of used predictors is limited to very small with small training set sizes (possibly causing lack of precision because important predictors are left out from the model), the model BPCR which uses all the 19 principal components results to predictions with better precision.

Estimated parameters for models BPCR and BPCRs for the case where all the field plots are used as the training set are shown in Table~\ref{table:fullparameter}. 
\begin{table}
\caption{Parameter estimates -- mean (95\% confidence interval with the median) -- for the forest inventory case with full data.\label{table:fullparameter}}
\begin{tabular}{l|c|c}\hline
 & BPCR & BPCRs \\\hline
 $\tau$ [Stem Number] & 354.1 (314.6 - 352.5 - 399.7) & 247.2 (205.3  - 245.9 -  290.3)\\
$\sigma$ [Stem Number] & - & 254.8 (198.4  - 251.5 -  290.3)\\
$\phi$ [m] & - & 535.5 (170.8  - 435.1 -  1484.6)\\
$\alpha_0$ & 1536.1 (427.2  - 1002.0 - 5883.3) & 1607.1 (433.5  - 992.7 -  6543.9)\\
$\alpha_1$ & 132.7 (82.8  - 127.5 -  214.3) & 130.6 (81.6 - 126.8 -  207.7)\\
\end{tabular}
\end{table}
The model BPCRs divides the total variability between spatial ($\sigma$) and non-spatial ($\tau$) and we see that for the BPCR $\tau^2$ is approximately equal to $\tau^2+\sigma^2$ for the BPCRs. This means that we get more accurate predictions when we are close enough to an observation location. Also, the table shows that the posterior distributions $\phi$ and $\alpha_0$ are skewed as mean values differ from the medians. By MCMC sampling for the predictive distributions, compared to using just a point estimate, we account the uncertainty in the parameters and the possible non-Gaussian features. The spatial correlation parameter $\phi$ in model BPCRs is about 500 m, but the range of possible values is large. Both models BPCR and BPCRs estimate the regularization parameters $\alpha_0$ and $\alpha_1$ similarly. 

\begin{figure}
  \centering
  \subfigure[BPCR: $\alpha_0$]{
  \includegraphics[width=0.48\textwidth]{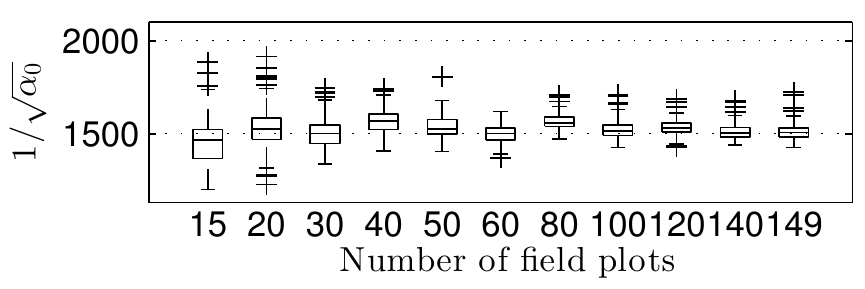}}
  \subfigure[BPCRs: $\alpha_0$]{
  \includegraphics[width=0.48\textwidth]{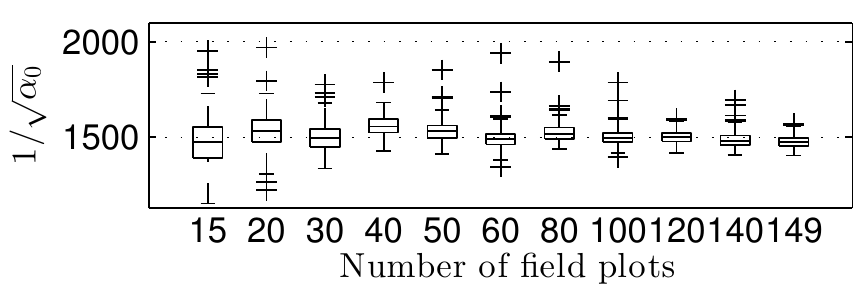}}
   \\
  \subfigure[BPCR: $\alpha_1$]{
  \includegraphics[width=0.48\textwidth]{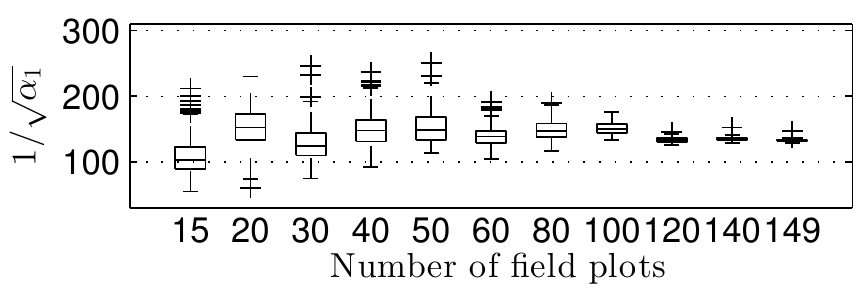}} 
  \subfigure[BPCRs: $\alpha_1$]{
  \includegraphics[width=0.48\textwidth]{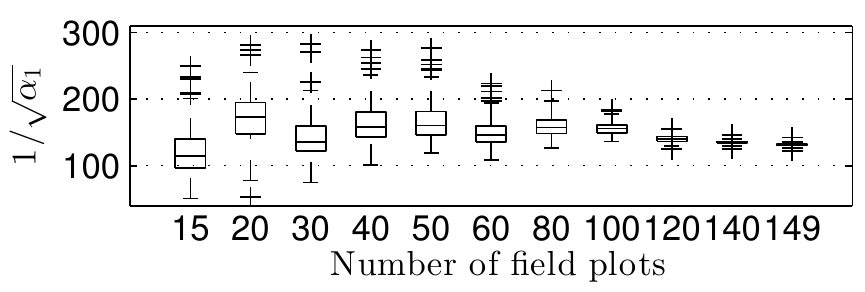}}
  \caption{Boxplot over 150 cross-validation experiments of the averages of estimated $1/\sqrt{\alpha}$-parameter values of models BPCR and BPCRs as a function of the number of field sample plots in the forest inventory case.   \label{fig:alpha_BPCR}}
\end{figure}

The hyper-prior parameters $\alpha_0$ and $\alpha_1$ for the inverse of the variance for the Gaussian prior regularize the maximum deviation from zero of the regression parameters $\beta$. The allowed standard deviation is  of size $\alpha_1^{-1/2}$ for parameters $\beta_j,\ j=1,2,\ldots,p$, and $\alpha_0^{-1/2}$ for the parameter $\beta_0$. The distribution of average estimated $\alpha_0$ and $\alpha_1$ values from the 150 LOO cross validations as a function of training set size $n$ are shown in Figure~\ref{fig:alpha_BPCR}.
The parameter values remain quite equal, however the variability is increasing when the training set size becomes smaller. With small training set size, the $\alpha$ values become larger, which equals to smaller allowed variance ($\alpha^{-1}$) and resulting smaller deviance from zero in the corresponding regression parameter value in vector $\vec{\beta}$.

\begin{figure}
  \centering
  \includegraphics[width=0.9\textwidth]{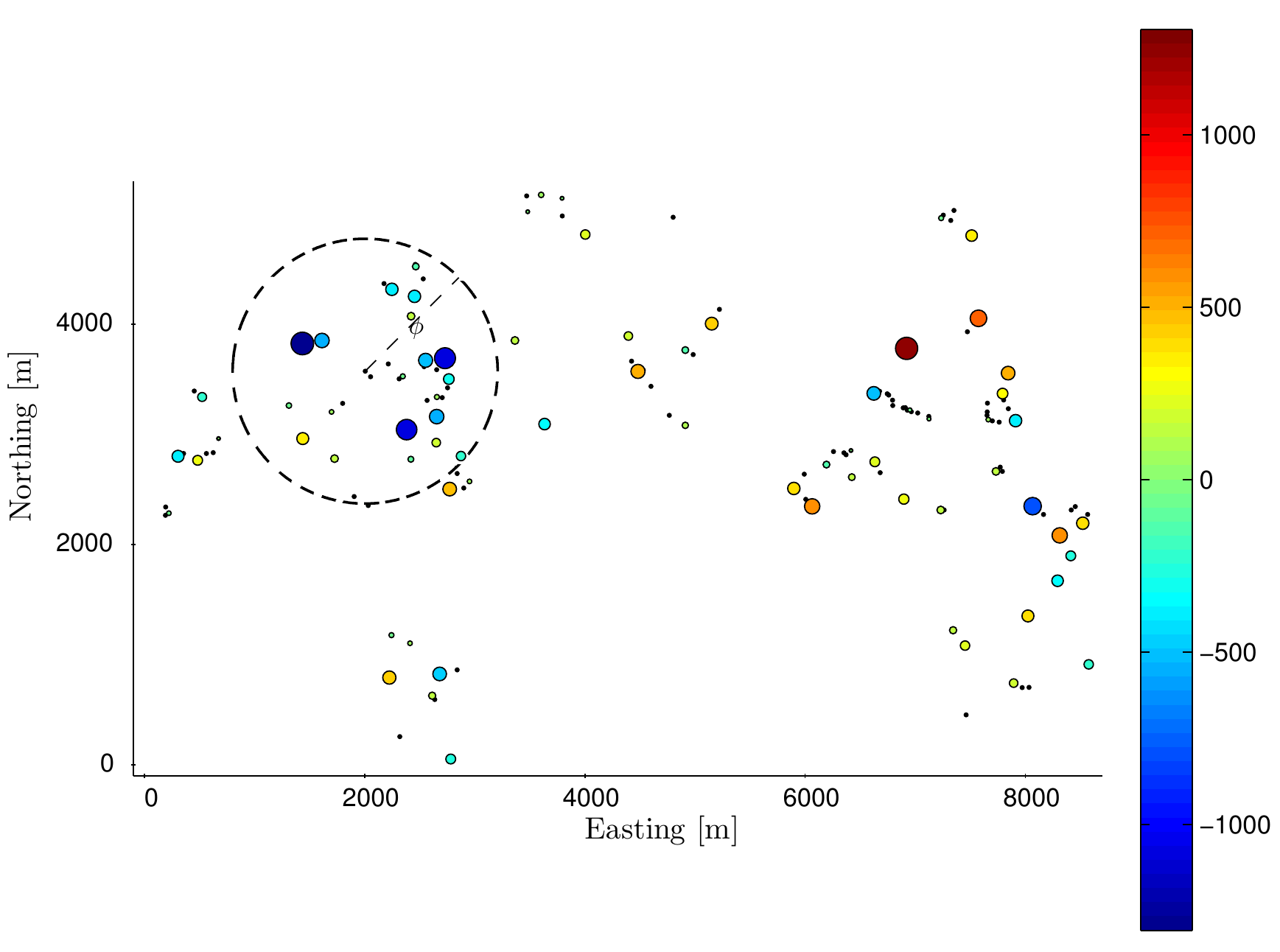}
  \caption{Example of spatial scatter of the residual estimated with the non-spatial BPCR model and $n = 80$ training set plots in the forest inventory case. The black dots show the plot locations of all the $N=150$ plots in the study area. The coloured circles show the relative magnitude of the residuals of the predictions in the training set plots. The size of the circles also show the absolute magnitude (i.e. deviance from zero) of the residual. Similar sized, similar coloured plots located close to each other indicate spatial correlation. The mean value of the estimated spatial range $\phi$ from the spatial model BPCRs is shown with dashed line around one field plot.   \label{fig:lieksa_residualmap}}
\end{figure}

Model BPCRs, spatial extension of the model BPCR, gives the best prediction precision. The model is similar to BPCR, except of the spatial, non-diagonal covariance structure defined by three parameters $\tau^2,\ \sigma^2$ and $\phi$. The residuals of the model BPCR show that part of the model error can be explained by spatial correlation among the residuals, see Figure~\ref{fig:lieksa_residualmap}, and this property is included in the model BPCRs.  

\begin{figure}
  \centering
  \subfigure[BPCR: $\tau^2$]{
  \includegraphics[width=0.48\textwidth]{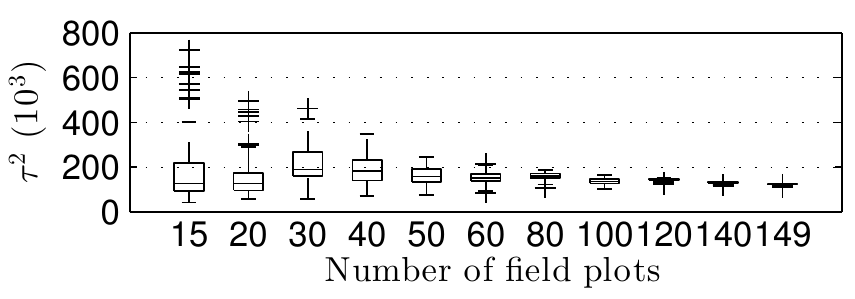}}  
  \subfigure[BPCRs: $\tau^2$]{
  \includegraphics[width=0.48\textwidth]{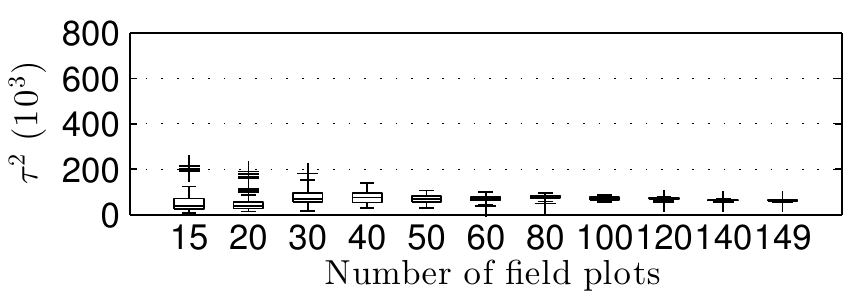}}\\
\subfigure[BPCRs: $\sigma^2$]{
  \includegraphics[width=0.48\textwidth]{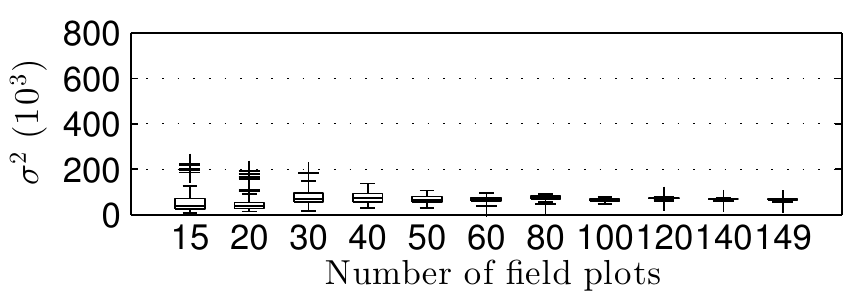}}
\subfigure[BPCRs: $\phi$]{
  \includegraphics[width=0.48\textwidth]{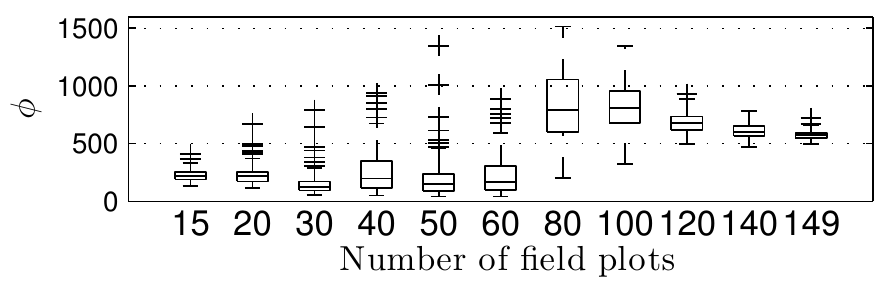}}  
  \caption{Boxplots of the estimated posterior mean variance/covariance parameter values for models BPCR and BPCRs as a function of the number of field sample plots and over the 150 leave-one-out cross-validation experiments. For non-spatial model BPCR, the parameter $\tau^2$ is the estimated residual variance.  \label{fig:lieksa_BPCRs}}
\end{figure}

The distribution of estimated mean values of variance/covariance parameters of models BPCR and BPCRs are shown as function of the training set size in Figure~\ref{fig:lieksa_BPCRs}. The values of average correlation length parameter $\phi$ correspond well with the phenomenon seen in Figure~\ref{fig:lieksa_res}. With about 80 and more field plots, the precision of the predictions made with the regularized model with spatial effects outperform the precision made without the spatial effects. With 80 plots, the average distance between the closest neighbouring field plots varies between 242 m and 272 m. To correctly estimate the parameter $\phi$, there must be enough data within the range of the effective distance of the spatial effect. Also, when using the model for prediction in a new location, the spatial effect helps only if the geographical distance from the nearest training plots is small compared to the estimated value of spatial range $\phi$. This is possible with about 80 or more field plots.

{A forest inventory case for prediction of a variable ``total volume'' for which the spatial correlation among model residuals is negligible is given in~\ref{sec:Aforestnospat}. }

\section{Conclusions and discussion}
\label{sec:discussion}

We proposed a Bayesian principal component regression model with spatial random effect, BPCRs. When the original predictors are highly correlated and when there is a spatial effect not explained by the fixed predictors of the linear model, BPCRs outperforms the other candidate models. It produced more accurate and precise predictions even with small training set sizes. In case there is no spatial correlation in the linear model residuals, the results are similar to those estimated with the same model without the spatial component, BPCR, and the analyses were free of artifacts described, e.g., by \citet{hodges10} and \citet{Paciorek10}. Overall, using Bayesian regularization with principal component regression makes the PCR analysis more robust and to have better predictive power compared to classical PCR analyses.

The computational cost for the model parameter estimation and for the prediction grows with the training set size as we need to operate on full covariance matrices. The training set size of approximately 150 made the computational time already quite high (several minutes for parameter estimation and prediction by MCMC). In typical forest inventory tasks the number of field plots might be several hundreds, and especially in case of large area inventory, the set of field sample plots should be relatively dense in order to capture the existing spatial effects in the model (the distances in the study area versus the location of field plots and the existing spatial correlation). 
When the training set size grows too large to be handled numerically by direct matrix operations, there are several alternatives available in the spatial statistics literature, such as reduced rank and multi-resolution methods and the use of different basis functions to represent the background model grid \citep{gelfand10,nguyen2014spatio,datta2016}.

We used standard exponential isotrophic correlation model for the spatial dependency. Even this brought significant benefits over the non-spatial models. By the same framework, it is possible to use more detailed information on the spatial structures, which might be available from separate studies. When we have spatial effects, one should carefully consider the best setting for the field plot locations to maximize the prediction precision. Also, it is important that the design of field measurement locations leads to unbiased predictions. The spatial effects can be estimated when they exists and the training set plot locations are correctly chosen.
To be useful, the locations for the predictions need to be within the effective spatial correlation distance of the training set locations also. The important study for the optimal designs is out of scope for this paper, but see \citet{junttila13}.

The proposed linear calibration model with spatial random effect and Bayesian regularization  allow for efficient sampling scheme that exploits the conditional structure of the hierarchical model. Only the three spatial parameters are sampled with a full Metropolis-Hastings MCMC algorithm, for the other parameters, conditional distributions are available for direct sampling. This allows for an efficient algorithm to account for the parameter uncertainty in the prediction and for full Bayesian uncertainty quantification for the predictions. The latter is crucially important in any real world environmental application.

\subsection*{Acknowledgements}
\label{sec:ack}
This research has been supported by the Academy of Finland INQUIRE project.

\bibliographystyle{elsarticle-harv}
\bibliography{spatial_btsvd}

\appendix

\section{Computational details}
\label{sec:Acomp}

Here, we give some further details on the computations outlined in Section~\ref{sec:comp}. Let $\mat{L}_\theta$ be a Cholesky decomposition of the covariance matrix $\vec{\Sigma}_\theta$, such that $\vec{\Sigma}_\theta = \mat{L}_\theta \mat{L}^T_\theta$. Define auxiliary variables $\tX_\theta$ and $\ty_\theta$ by scaling the design matrix $\mat{X}$ and observation vector $\vec{y}$ by $\mat{L}_\theta$ and augmenting  with rows from the prior as
\begin{equation}
  \tX_\theta =
  \begin{bmatrix}
\mat{L}^{-1}_\theta \mat{X} \\ \vec{\alpha} \mat{I}_p,
  \end{bmatrix},\qquad
\ty_\theta =
\begin{bmatrix}
 \mat{L}^{-1}_\theta \vec{y}\\ \vec{0}_p
\end{bmatrix}.
\end{equation}
Next, define sum of squares as
\begin{equation}
  \label{eq:ss}
  SS_\beta = (\ty_\theta-\tX_\theta\vec{\beta})^T(\ty_\theta-\tX_\theta\vec{\beta})
\end{equation}
and conditional mean and covariance estimates as
\begin{align}
  \hb &= (\tX_\theta^T\tX_\theta)^{-1}\tX_\theta^T\ty_\theta,   \label{eq:betahat}\\
  \widehat{\vec{\Sigma}} &= (\tX_\theta^T\tX_\theta)^{-1}.  \label{eq:Shat}
\end{align}

Now, we can sample $\vec{\beta}|\vec{y},\vec{\alpha},\vec{\theta}$ directly from the multivariate Gaussian distribution $N(\hb,\widehat{\vec{\Sigma}})$.
Then, after updating the value of $\vec{\beta}$, we can sample $\vec{\alpha}|\vec{y},\vec{\beta},\vec{\theta}$ (for each $\alpha_i\in\vec{\alpha}$) from 
\begin{equation}
  \label{eq:alphacond}
  {\alpha}_i|\vec{y},\vec{\beta},\vec{\theta} \sim \chisq\left(n + \nu_i, \frac{{\nu_i} {\alpha_i}^{-1} + SS_\beta}{n + \nu_i}\right).
\end{equation}

For $\vec{\theta}|\vec{y},\vec{\beta},\vec{\alpha}$, the conditional log posterior for $\vec{\theta}$, given $\vec{\beta}$ and $\vec{\alpha}$ can be written as
\begin{equation}
  \label{eq:thetalik}
 \log p(\vec{\theta}|\vec{y},\vec{\beta},\vec{\alpha}) = \mathrm{c}  -\frac12  (\vec{y}-\mat{X}\vec{\beta})^T\vec{\Sigma}^{-1}_\theta(\vec{y}-\mat{X}\vec{\beta})
- \log(|\vec{\Sigma}_\theta|) - \frac12 \sum_{j=1}^{3}\left(\frac{\log(\theta_j/\mu_{\theta j})}{\sigma_{\theta j}}\right)^2,
\end{equation}
where $\mathrm{c}$ is a constant that does not depend on $\vec{\theta}$ and the last sum term in equation~(\ref{eq:thetalik}) comes from the log-normal prior for $\vec{\theta}$ in equation~(\ref{equ:thetapri}).

The Metropolis-Hastings step for updating the value of $\vec{\theta}$ is the following. Given the current value of $\vec{\theta}$, propose a new candidate $\vec{\theta}^*$ from a 3-dimensional Gaussian proposal distribution. Using the likelihood from equation (\ref{eq:thetalik}), the value is accepted with probability
\begin{equation}
  \label{eq:accep}
  \mathrm{min}\left(1 , \frac{p(\vec{\theta}^*|\vec{y},\vec{\beta},\vec{\alpha})}{p(\vec{\theta}|\vec{y},\vec{\beta},\vec{\alpha})}\right).
\end{equation}
If not accepted, $\vec{\theta}$ is kept at the previous value in the next iteration of updating $\vec{\beta}$ and $\vec{\alpha}$. As the acceptance probability in equation (\ref{eq:accep}) depends on the ratio of the conditional distribution evaluated at two parameter values, the unknown constant $c$ in equation (\ref{eq:thetalik}) cancels out. This Metropolis-Hastings step can be used with adaptive tuning of the proposal covariance matrix \citep{dram} and the whole sampling procedure is repeated until the MCMC chain has converged and enough values have been sampled. Convergence of the algorithm can be assessed by calculating convergence diagnostics and studying chain plots. The adequate chain length is determined by estimating the effective sample size and the Monte Carlo error of the chain-based estimates.

\section{Synthetic data generation}
\label{sec:Asimu}

We describe the procedure to generate the synthetic data used in Section~\ref{test:toy}.
To simulate a system with highly correlated predictors, nine correlated, normally distributed random predictors with condition number 30, $\mat{Z}_\mathrm{corr}$, and five independent, normally distributed predictors, $\mat{Z}_\mathrm{noise}$, are generated. To mix the predictors, orthonormal rotation, $\mat{P}_\mathrm{rot}$, is performed. The complete set of correlated predictors contain the intercept and the rotated predictors as
\begin{equation}
  \mat{Z} =
  \begin{bmatrix}
    \vec{1} &  
    \begin{bmatrix}
      \mat{Z}_\mathrm{corr}  &\mat{Z}_\mathrm{noise}
    \end{bmatrix} \mat{P}_\mathrm{rot}\\
  \end{bmatrix}.
\end{equation}
The covariance matrix for the correlated predictors is constructed with singular values decreasing from one to $1/\mathrm{(condition\ number)}$ in such a way that the inverses are linearly increasing. The first singular vector is assigned randomly.

Arbitrarily selected intercept, $\beta_\mathrm{int}=20$, and the first five correlated predictors, $\beta_\mathrm{reg} = \left[ 1, 2, 3, 2, 1\right]$, affect the response by the linear model. The other predictors are discarded by setting their regression parameters to zero and the full true parameter vector will be
\begin{equation}
  \vec{\beta} =
  \begin{bmatrix}
    \beta_\mathrm{int} &  
    \begin{bmatrix}
      \vec{\beta}_\mathrm{reg}  &\vec{0}
    \end{bmatrix} \mat{P}_\mathrm{rot}\\
  \end{bmatrix}^{T}.
\end{equation}

\section{Forest inventory, total volume prediction}
\label{sec:Aforestnospat}

\setcounter{figure}{0}
\setcounter{table}{0}

In addition to the ecological variable stem number, also the variable total volume was predicted using the same set of validation models and procedure given in Section~\ref{test:forest}. We didn't detect any spatial correlation among the model residuals for this variable, thus the corresponding models with and without spatial effects, e.g. BPCR and BPCRs, should result to similar performance. 

\begin{figure}[h]
  \centering
  \subfigure[Models PCR and PCRs.]{
  \includegraphics[width=0.48\textwidth]{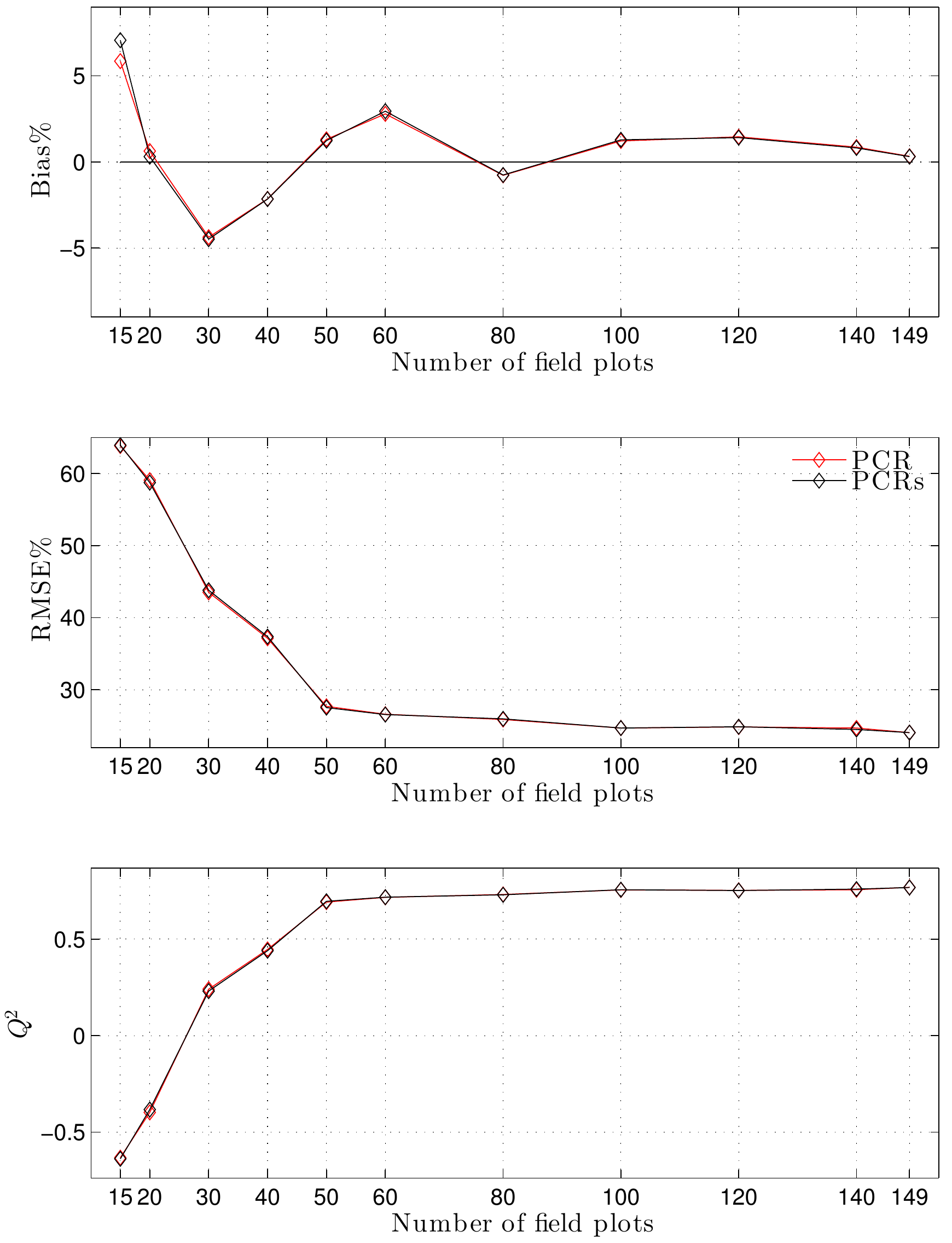}} 
  \subfigure[Models PCR(s) adapt and BPCR(s).]{
  \includegraphics[width=0.48\textwidth]{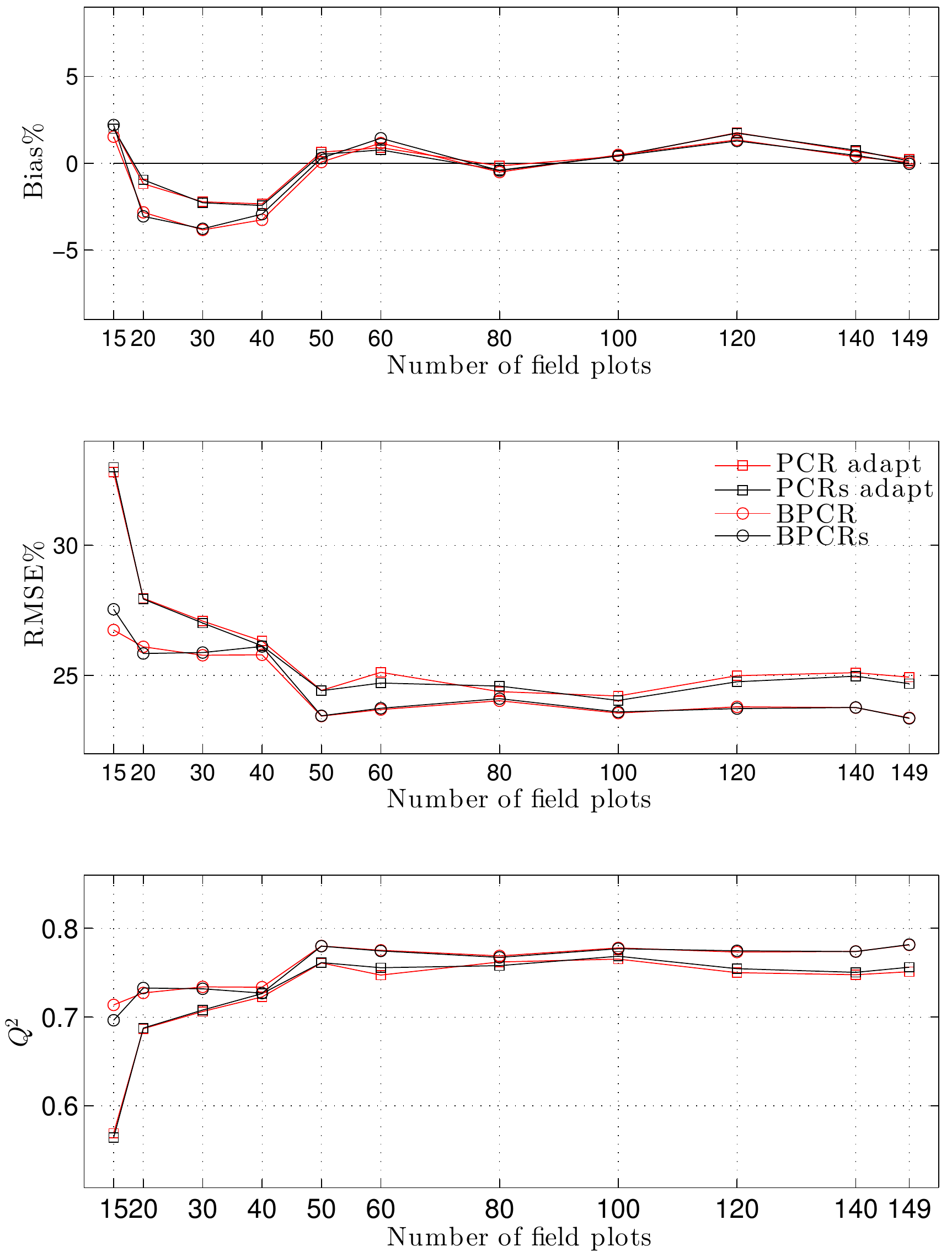}} 
  \caption{Bias\%, RMSE\% and $Q^2$ for different models in the forest inventory case with forest attribute total volume.  \label{fig:lieksa_res_nospat}}
\end{figure}

For the total volume, prediction performance of the six different models, PCR,  PCRs, PCR adapt, PCRs adapt, BPCR and BPCRs, are evaluated similar manner as those for the Stem number in Section~\ref{sec:forinv_stemno}.
The baseline results for the case with full training set size are given in Table~\ref{tab:loo_sols_nospat}.
\begin{table}[h]
\caption{Error statistics of the Leave-One-Out models ($n=149$) in the forest inventory case with ecological variable total volume.\label{tab:loo_sols_nospat}}
\centering
\begin{tabular}{l| c c| c c| c c}
 & PCR	 & PCRs	 & PCR adapt	 & PCRs adapt	 & BPCR	 & BPCRs	 \\\hline
 Bias\% & 0.33 	 & 0.31 	 & 0.14 	 & 0.24 	 & 0.06 	 & -0.03 	 \\
 RMSE\% & 24.05 	 & 24.06 	 & 24.68 	 & 24.93 	 & 23.37 	 & 23.35 	 \\
 $Q^2$ & 0.77 	 & 0.77 	 & 0.76 	 & 0.75 	 & 0.78 	 & 0.78 	 \\
\end{tabular}
\end{table}
The results show that for total volume, the models BPCR and BPCRs give the best predictions, although the differences among all the results are small. The differences among models with and without spatial effects are even smaller than the differences between model types. The same effect can be seen when reducing the training set size, see Figure~\ref{fig:lieksa_res_nospat}. The proposed model, together with its version without spatial effects, outperforms the other models, and the difference among the models with and without spatial effects of the same model type are negligible.

\end{document}